\documentclass[twocolumn]{aastex701}
\usepackage{tablefootnote}
\usepackage{hyperref}
\usepackage{amsmath}
\usepackage{color}
\usepackage{upgreek}
\usepackage{xspace}
\usepackage{tipa}

\begin{document}

\title{A steadily declining dispersion measure for the repeating fast radio burst FRB~20220529A:\\ Evidence for an FRB engine embedded in an expanding supernova remnant}
\shorttitle{The dynamic source environment of FRB~20220529A}

\author[0000-0002-8897-1973]{Ayush Pandhi}
\email{ayush.pandhi@mcgill.ca}
\affiliation{Department of Physics, McGill University, 3600 rue University, Montr\'eal, QC H3A 2T8, Canada}
\affiliation{Trottier Space Institute, McGill University, 3550 rue University, Montr\'eal, QC H3A 2A7, Canada}
\affiliation{David A.\ Dunlap Department of Astronomy and Astrophysics, 50 St. George Street, University of Toronto, ON M5S 3H4, Canada}
\affiliation{Dunlap Institute for Astronomy and Astrophysics, 50 St. George Street, University of Toronto, ON M5S 3H4, Canada}
\correspondingauthor{Ayush Pandhi}

\author[0000-0003-0510-0740]{Kenzie Nimmo}
\email{knimmo@northwestern.edu}
\altaffiliation{NHFP Einstein Fellow}
\affiliation{Center for Interdisciplinary Exploration and Research in Astronomy, Northwestern University, 1800 Sherman Avenue, Evanston, IL 60201, USA}
\affiliation{MIT Kavli Institute for Astrophysics and Space Research, Massachusetts Institute of Technology, 77 Massachusetts Ave, Cambridge, MA 02139, USA}

\author[0000-0002-3980-815X]{Shion Andrew}
\email{shiona@mit.edu}
\affiliation{MIT Kavli Institute for Astrophysics and Space Research, Massachusetts Institute of Technology, 77 Massachusetts Ave, Cambridge, MA 02139, USA}
\affiliation{Department of Physics, Massachusetts Institute of Technology, 77 Massachusetts Ave, Cambridge, MA 02139, USA}

\author[0000-0002-1800-8233]{Charanjot Brar}
\email{charanjot.brar@nrc-cnrc.gc.ca}
\affiliation{NRC Herzberg Astronomy and Astrophysics, 5071 West Saanich Road, Victoria, BC V9E2E7, Canada}

\author[0000-0002-2878-1502]{Shami Chatterjee}
\email{shami@astro.cornell.edu}
\affiliation{Cornell Center for Astrophysics and Planetary Science, Cornell University, Ithaca, NY 14853, USA}

\author[0000-0001-6422-8125]{Amanda M.~Cook}
\email{amanda.cook@mail.mcgill.ca}
\affiliation{Department of Physics, McGill University, 3600 rue University, Montr\'eal, QC H3A 2T8, Canada}
\affiliation{Trottier Space Institute, McGill University, 3550 rue University, Montr\'eal, QC H3A 2A7, Canada}
\affiliation{Anton Pannekoek Institute for Astronomy, University of Amsterdam, Science Park 904, 1098 XH Amsterdam, The Netherlands}

\author[0000-0002-8376-1563]{Alice Curtin}
\email{alice.curtin@mail.mcgill.ca}
\affiliation{Department of Physics, McGill University, 3600 rue University, Montr\'eal, QC H3A 2T8, Canada}
\affiliation{Trottier Space Institute, McGill University, 3550 rue University, Montr\'eal, QC H3A 2A7, Canada}
\affiliation{Anton Pannekoek Institute for Astronomy, University of Amsterdam, Science Park 904, 1098 XH Amsterdam, The Netherlands}

\author[0000-0002-3382-9558]{B.~M.~Gaensler}
\email{gaensler@ucsc.edu}
\affiliation{Department of Astronomy and Astrophysics, University of California, Santa Cruz, 1156 High Street, Santa Cruz, CA 95060, USA}
\affiliation{Dunlap Institute for Astronomy and Astrophysics, 50 St. George Street, University of Toronto, ON M5S 3H4, Canada}
\affiliation{David A.\ Dunlap Department of Astronomy and Astrophysics, 50 St. George Street, University of Toronto, ON M5S 3H4, Canada}

\author[0000-0003-4056-4903]{Marcin Gawroński}
\email{motylek@astro.uni.torun.pl}
\affiliation{Institute of Astronomy, Faculty of Physics, Astronomy and Informatics, Nicolaus Copernicus University, Grudziadzka 5, 87-100 Toruń, Poland}

\author[0000-0003-2317-1446]{Jason Hessels}
\email{jason.hessels@mcgill.ca}
\affiliation{Department of Physics, McGill University, 3600 rue University, Montr\'eal, QC H3A 2T8, Canada}
\affiliation{Trottier Space Institute, McGill University, 3550 rue University, Montr\'eal, QC H3A 2A7, Canada}
\affiliation{Anton Pannekoek Institute for Astronomy, University of Amsterdam, Science Park 904, 1098 XH Amsterdam, The Netherlands}
\affiliation{ASTRON, Netherlands Institute for Radio Astronomy, Oude Hoogeveensedijk 4, 7991 PD Dwingeloo, The Netherlands}

\author[0000-0001-9345-0307]{Victoria M.~Kaspi}
\email{victoria.kaspi@mcgill.ca}
\affiliation{Department of Physics, McGill University, 3600 rue University, Montr\'eal, QC H3A 2T8, Canada}
\affiliation{Trottier Space Institute, McGill University, 3550 rue University, Montr\'eal, QC H3A 2A7, Canada}
\affiliation{Lowy Distinguished Guest Professor, Tel Aviv University, Tel Aviv, Israel}

\author[0009-0004-4176-0062]{Afrokk Khan}
\email{afrasiyab.khan@mcgill.ca}
\affiliation{Department of Physics, McGill University, 3600 rue University, Montr\'eal, QC H3A 2T8, Canada}
\affiliation{Trottier Space Institute, McGill University, 3550 rue University, Montr\'eal, QC H3A 2A7, Canada}

\author[0000-0001-6664-8668]{Franz Kirsten}
\email{franz.kirsten@chalmers.se}
\affiliation{Department of Space, Earth and Environment, Chalmers University of Technology, Onsala Space Observatory, 439 92, Onsala, Sweden}
\affiliation{ASTRON, Netherlands Institute for Radio Astronomy, Oude Hoogeveensedijk 4, 7991 PD Dwingeloo, The Netherlands}

\author[0000-0002-5857-4264]{Mattias Lazda}
\email{mattias.lazda@mail.utoronto.ca}
\affiliation{Dunlap Institute for Astronomy and Astrophysics, 50 St. George Street, University of Toronto, ON M5S 3H4, Canada}
\affiliation{David A.\ Dunlap Department of Astronomy and Astrophysics, 50 St. George Street, University of Toronto, ON M5S 3H4, Canada}

\author[0000-0002-4209-7408]{Calvin Leung}
\email{calvin_leung@berkeley.edu}
\affiliation{Miller Institute for Basic Research, Stanley Hall, Room 206B, Berkeley, CA 94720, USA}
\affiliation{Department of Astronomy, University of California, Berkeley, CA 94720, United States}

\author[0000-0002-7164-9507]{Robert Main}
\email{robert.main@mcgill.ca}
\affiliation{Department of Physics, McGill University, 3600 rue University, Montr\'eal, QC H3A 2T8, Canada}
\affiliation{Trottier Space Institute, McGill University, 3550 rue University, Montr\'eal, QC H3A 2A7, Canada}

\author[0000-0002-4279-6946]{Kiyoshi W.~Masui}
\email{kmasui@mit.edu}
\affiliation{MIT Kavli Institute for Astrophysics and Space Research, Massachusetts Institute of Technology, 77 Massachusetts Ave, Cambridge, MA 02139, USA}
\affiliation{Department of Physics, Massachusetts Institute of Technology, 77 Massachusetts Ave, Cambridge, MA 02139, USA}

\author[0000-0001-7348-6900]{Ryan Mckinven}
\email{ryan.mckinven@mcgill.ca}
\affiliation{Department of Physics, McGill University, 3600 rue University, Montr\'eal, QC H3A 2T8, Canada}
\affiliation{Trottier Space Institute, McGill University, 3550 rue University, Montr\'eal, QC H3A 2A7, Canada}

\author[0000-0002-2551-7554]{Daniele Michilli}
\email{danielemichilli@gmail.com}
\affiliation{Laboratoire d'Astrophysique de Marseille, Aix-Marseille Univ., CNRS, CNES, Marseille, France}

\author[0000-0002-0940-6563]{Mason Ng}
\email{mason.ng@mcgill.ca}
\affiliation{Department of Physics, McGill University, 3600 rue University, Montr\'eal, QC H3A 2T8, Canada}
\affiliation{Trottier Space Institute, McGill University, 3550 rue University, Montr\'eal, QC H3A 2A7, Canada}

\author[0000-0001-9381-8466]{Omar Ould-Boukattine}
\email{ouldboukattine@astron.nl}
\affiliation{ASTRON, Netherlands Institute for Radio Astronomy, Oude Hoogeveensedijk 4, 7991 PD Dwingeloo, The Netherlands}
\affiliation{Anton Pannekoek Institute for Astronomy, University of Amsterdam, Science Park 904, 1098 XH Amsterdam, The Netherlands}

\author[0000-0002-8912-0732]{Aaron B.~Pearlman}
\email{aaron.b.pearlman@mit.edu}
\altaffiliation{NASA Hubble Fellow}
\affiliation{MIT Kavli Institute for Astrophysics and Space Research, Massachusetts Institute of Technology, 77 Massachusetts Ave, Cambridge, MA 02139, USA}
\affiliation{Department of Physics, Massachusetts Institute of Technology, 77 Massachusetts Ave, Cambridge, MA 02139, USA}
\affiliation{Department of Physics, McGill University, 3600 rue University, Montr\'eal, QC H3A 2T8, Canada}
\affiliation{Trottier Space Institute, McGill University, 3550 rue University, Montr\'eal, QC H3A 2A7, Canada}

\author[0000-0002-4795-697X]{Ziggy Pleunis}
\email{z.pleunis@uva.nl}
\affiliation{Anton Pannekoek Institute for Astronomy, University of Amsterdam, Science Park 904, 1098 XH Amsterdam, The Netherlands}
\affiliation{ASTRON, Netherlands Institute for Radio Astronomy, Oude Hoogeveensedijk 4, 7991 PD Dwingeloo, The Netherlands}

\author[0000-0002-3430-7671]{Alexander W.~Pollak}
\email{apollak@seti.org}
\affiliation{SETI Institute, 339 Bernardo Avenue, Suite 200 Mountain View, CA 94043, USA}

\author[0009-0008-2000-6959]{Sachin Pradeep~E.~T.}
\email{sachin.pradeepetakkepravanthulicheri@mail.mcgill.ca}
\affiliation{Department of Physics, McGill University, 3600 rue University, Montr\'eal, QC H3A 2T8, Canada}
\affiliation{Trottier Space Institute, McGill University, 3550 rue University, Montr\'eal, QC H3A 2A7, Canada}

\author[0000-0003-2422-6605]{Weronika Puchalska}
\email{wpuchalska@doktorant.umk.pl}
\affiliation{Institute of Astronomy, Faculty of Physics, Astronomy and Informatics, Nicolaus Copernicus University, Grudziadzka 5, 87-100 Toruń, Poland}

\author[0000-0002-4623-5329]{Mawson W.~Sammons}
\email{mawson.sammons@mcgill.ca}
\affiliation{Department of Physics, McGill University, 3600 rue University, Montr\'eal, QC H3A 2T8, Canada}
\affiliation{Trottier Space Institute, McGill University, 3550 rue University, Montr\'eal, QC H3A 2A7, Canada}

\author[0000-0002-7374-7119]{Paul Scholz}
\email{pscholz@yorku.ca}
\affiliation{Department of Physics and Astronomy, York University, 4700 Keele Street, Toronto, ON MJ3 1P3, Canada}

\author[0000-0002-4823-1946]{Vishwangi Shah}
\email{vishwangi.shah@mail.mcgill.ca}
\affiliation{Department of Physics, McGill University, 3600 rue University, Montr\'eal, QC H3A 2T8, Canada}
\affiliation{Trottier Space Institute, McGill University, 3550 rue University, Montr\'eal, QC H3A 2A7, Canada}

\author[0000-0002-6823-2073]{Kaitlyn Shin}
\email{kaitshin@caltech.edu}
\affiliation{Division of Physics, Mathematics, and Astronomy, California Institute of Technology, Pasadena, CA 91125, USA}

\author[0000-0003-2631-6217]{Seth R.~Siegel}
\email{ssiegel@perimeterinstitute.ca}
\affiliation{Perimeter Institute of Theoretical Physics, 31 Caroline Street North, Waterloo, ON N2L 2Y5, Canada}
\affiliation{Department of Physics, McGill University, 3600 rue University, Montr\'eal, QC H3A 2T8, Canada}
\affiliation{Trottier Space Institute, McGill University, 3550 rue University, Montr\'eal, QC H3A 2A7, Canada}

\author[0000-0002-2088-3125]{Kendrick Smith}
\email{kmsmith@perimeterinstitute.ca}
\affiliation{Perimeter Institute of Theoretical Physics, 31 Caroline Street North, Waterloo, ON N2L 2Y5, Canada}

\begin{abstract} 
We present the discovery and subsequent 3.2~year monitoring campaign of the repeating fast radio burst FRB~20220529A with CHIME/FRB. We observe a gradual dispersion measure (DM) decline of $-0.881\pm0.001~\mathrm{pc}~\mathrm{cm}^{-3}~\mathrm{year}^{-1}$ ($-1.235\pm0.001~\mathrm{pc}~\mathrm{cm}^{-3}~\mathrm{year}^{-1}$ in the rest frame), implying a $\geq3.5\pm0.2$\% decrease of the total electron column in the source environment, and we see scattering timescale variations over weeks to years. We observe a short-lived excursion in which the DM rises by $\sim 1~\mathrm{pc}~\mathrm{cm}^{-3}$, immediately preceding a transient $\sim 2000~\mathrm{rad}~\mathrm{m}^{-2}$ Faraday rotation measure (RM) increase previously reported for this source, before returning to its gradual DM decline. We identify a local line-of-sight magnetic field around FRB~20220529A during this DM/RM excursion of $3.4 \pm 0.2~\mathrm{mG}$, corresponding to one of the most strongly magnetized FRB environments. We measure a decrease in the linear polarization fraction of FRB~20220529A bursts with decreasing frequency that we attribute to depolarization from multi-path propagation in the source environment. We also place a $5\sigma$ upper limit on the spectral luminosity of an associated persistent radio source of $\leq 5\times10^{28}~\mathrm{erg}~\mathrm{s}^{-1}~\mathrm{Hz}^{-1}$ at 1.5~GHz. These observations are consistent with FRB~20220529A originating from a young ($\sim$~years to centuries old) expanding supernova remnant, with short-lived DM and RM variability arising from interactions with the supernova remnant or with a binary companion.
\end{abstract}


\section{Introduction} \label{sec:intro}
Fast radio bursts (FRBs) are micro- to millisecond duration radio transients with (mostly) extragalactic origins \citep{2007Sci...318..777L, 2022A&ARv..30....2P}. While one FRB source has been linked to a Galactic magnetar \citep{2020Natur.587...54C, 2020Natur.587...59B}, the specific origins of most FRBs remain unresolved. Approximately $3\%$ of FRB sources have been seen to repeat (i.e., they produce more than one burst over their lifetime; Cook et al. {\it in prep}). This behavior has fueled developments in our understanding of repeating FRBs, through follow-up observations to localize repeaters to their source environments \citep[e.g.,][]{2017Natur.541...58C, 2020Natur.577..190M, 2022Natur.602..585K,Moroianu_2026_ApJL}, to understand their energy distributions \citep[e.g.,][]{2021Natur.598..267L, 2024NatAs...8..337K, ouldboukattine_2025_mnras}, to study their complex morphology over a range of frequencies \citep[e.g.,][]{2018ApJ...863....2G, 2019ApJ...876L..23H, 2021ApJ...911L...3P, 2022NatAs...6..393N, 2025ApJ...992..206C}, and to constrain multi-wavelength counterparts \citep[e.g.,][]{2023ApJ...947L..28H, 2024ApJ...974..170C, 2025NatAs...9..111P,Gouiffes_2025_A&A}. The propagation effects that are imparted on the signal from ionized, magnetized, and turbulent media that the FRB encounters can be monitored over the history of repeating sources, informing us of the dynamics of their source environments \citep[e.g.,][]{2023MNRAS.519..821O, 2025ApJ...982..154N, 2025arXiv251011352S}.

FRB signals are modulated by a number of frequency-dependent propagation effects as they pass through ionized and magnetized plasma along the line of sight (LOS). Dispersion, quantified by the dispersion measure (DM), traces the integrated electron column density toward the source. Multi-path propagation through inhomogeneous plasma produces temporal broadening of the burst, with the scattering timescale ($\tau$) probing the amplitude and scale of electron density fluctuations in the scattering media. As most FRBs exhibit significant levels of linear polarization \citep[e.g.,][]{2024ApJ...968...50P, 2024ApJ...964..131S, 2025PASA...42..133S}, their signals undergo Faraday rotation, characterized by the rotation measure (RM), which traces the product of electron density and LOS magnetic field strength. These observables represent a superposition of contributions from multiple plasma components (e.g., the Milky Way and host galaxy interstellar media and halo, circumgalactic media, the intergalactic medium). However, a dynamic environment surrounding the FRB source can lead to measurable changes in the amplitude of these propagation effects over timescales of variation of days to years.

Secular DM changes on timescales of months or longer have only been observed for a small number of repeating FRB sources \citep[e.g.,][]{2023MNRAS.526.3652K, 2025arXiv250715790W, 2025arXiv251011352S, 2025arXiv250916374O, NIU202676}. FRB~20190520B exhibits changes in the scattering timescale between bursts separated by only $\sim3$~minutes \citep{2023MNRAS.519..821O}, suggesting the presence of an inhomogeneous and turbulent source environment. This is also consistent with possible DM variations in FRB~20190520B \citep{2023Sci...380..599A}, though it remains unclear whether these represent true DM changes or are instead caused by unresolved temporal structure. Spectral depolarization reported in some repeating FRBs is consistent with multi-path propagation through an inhomogeneous magnetoionic environment local to the FRB source \citep{2022Sci...375.1266F}. Compared to DM and $\tau$, RM variations in repeating FRBs are more commonly observed \citep[e.g.,][]{2018Natur.553..182M,2021ApJ...908L..10H, Xu_2022_Natur,2023ApJ...951...82M, 2023ApJ...956...23S,2024MNRAS.527.9872G,2025ApJ...982..154N} and they depict many distinct trends, such as: linear evolution \citep{2023ApJ...950...12M, 2026arXiv260216409U}, LOS magnetic field reversals \citep{2023Sci...380..599A}, and oscillating variations \citep{2022NatCo..13.4382W}.

After the discovery of the repeating source FRB~20220529A by the CHIME/FRB Collaboration via a Virtual Observatory Event (VOEvent) alert \citep{2025AJ....169...39A}, follow-up observations with the Five-hundred-meter Aperture Spherical Telescope (FAST) revealed a rapid RM increase of $\sim2000~\mathrm{rad~m^{-2}}$ on a timescale of $\lesssim2$ months, followed by a decline to $|\mathrm{RM}|\sim10$–$100~\mathrm{rad~m^{-2}}$ within two weeks \citep{2026Sci...391..280L}. A plausible physical interpretation presented by \cite{2026Sci...391..280L} of this RM excursion is a coronal mass ejection (CME) generated by a hypothesized binary companion to the FRB source that intersects the propagation path of the FRB emission. Other possible explanations include: small-scale turbulence from a surrounding supernova remnant or pulsar wind nebula \citep{2026Sci...391..280L}, orbital motion between a magnetar and stellar binary companion \citep{2026arXiv260102734D, 2026Sci...391..280L}, or interactions between magnetar ejecta shells with different velocities \citep{2025A&A...698L...3X}.

In this paper, we present the discovery of 16 FRB~20220529A bursts detected with CHIME/FRB ($400$–$800$~MHz) over $\sim3.2$ years of monitoring. We also report an additional burst detected at 1.4~GHz with the Westerbork telescope as part of a Hyperflash campaign spanning $\sim2.5$~years. In Section~\ref{sec:obs}, we describe the observations and analysis methods used to characterize propagation effects. Using our low-frequency CHIME/FRB data, we precisely track the long-term evolution of the DM, RM and $\tau$, and find a steady, secular decline in the DM over the multi-year observing campaign. These results, together with an investigation of the frequency-dependent depolarization of FRB~20220529A, and a search for a persistent radio source (PRS) at the FRB position, are presented in Section~\ref{sec:results}. In Section~\ref{sec:discussion}, we interpret our measurements within a framework in which the FRB source is a compact object embedded in an expanding supernova remnant, which we find to be the most likely scenario. We also test various models of FRB source environments that may cause a short-lived excursion in the observed DM and RM, before summarizing our conclusions in Section~\ref{sec:conclusions}.

\section{Observations and methods} \label{sec:obs}
\subsection{CHIME/FRB} \label{sec:obs_chime}
CHIME is a radio telescope operating between $400-800$~MHz that is hosted by the Dominion Radio Astrophysical Observatory in British Columbia, Canada. CHIME observes at declinations $\gtrsim -11^{\circ}$ on a daily cadence, making it proficient at discovering new FRBs and monitoring repeating sources. There are two data streams for CHIME/FRB: (i) for FRBs with a signal to noise ratio S/N $> 8$, total intensity data with a time resolution of $0.983$~ms and frequency resolution of $24.4$~kHz are saved; (ii) for FRBs with $\mathrm{S/N} > 12$,\footnote{For bursts identified as originating from a known repeater, the S/N threshold is decreased to $\mathrm{S/N} >10$ for baseband data to be saved to disk.} raw voltage data (i.e., ``baseband data'') with full polarization are also recorded at a time resolution of $2.56~\mu\mathrm{s}$ and with a frequency resolution of $390.625$~kHz \citep{2021ApJS..257...59C, Michilli2024}.

FRB~20220529A was discovered by CHIME/FRB on 2022 May 29 with a real-time detection $\mathrm{S/N} = 16.8$,\footnote{The real-time CHIME/FRB search pipeline is described in detail by \cite{2018ApJ...863...48C}} and it is statistically classified as a repeating FRB by Cook et al. ({\it in prep}). In all, CHIME/FRB has detected a total of 26 bursts from this source (16 with baseband data), with the most recent detection on 2025 August 11. Due to its low declination ($\sim20^{\circ}$), the average exposure time of CHIME/FRB at the position of this source is approximately $2.7$~minutes per day \citep{2021ApJS..257...59C}. Both the initial burst and subsequent repeat bursts were broadcast to the community through the VOEvents service \citep{2025AJ....169...39A}. The $2.56~\mu\mathrm{s}$ resolution baseband data for FRB~20220529A show that almost all of its bursts have complex morphology, often with multiple components and downward drifting structure in frequency (Figure~\ref{fig:waterfalls}). Therefore, we only measure propagation effects (DM, RM, and $\tau$) in the bursts with baseband data, as intensity data do not provide polarization information, and their coarse time resolution makes it difficult to disentangle DM and $\tau$ from downward-drifting structures \citep[e.g., see][]{2025ApJ...979..160S}. First the baseband data are beamformed \citep{2021ApJ...910..147M} to the Karl~G.~Jansky Very Large Array (VLA) \textit{realfast} localization presented in \cite{2026Sci...391..280L}. We independently localize FRB~20220529A using data that we simultaneously recorded with the CHIME/FRB Outriggers \citep{2025ApJ...993...55C} for a subset of our CHIME/FRB bursts, and find our position is in agreement with the \textit{realfast} localization (see Appendix \ref{sec:appendix_A}). The data are coherently dedispersed to a S/N-maximized DM initially, and then the DM is further refined using the {\tt DM\_PHASE}\footnote{\url{https://github.com/danielemichilli/DM_phase}} package, which measures a structure-optimizing DM \citep{2019ascl.soft10004S}. 

We then fit the dedispersed Stokes I dynamic spectra using {\tt fitburst} \citep{2024ApJS..271...49F}, which uses a least-squares optimization routine to estimate the time of arrival (TOA) as a Modified Julian Date (MJD; topocentric at CHIME and referenced to 400~MHz), pulse width ($w$) in ms, scattering timescale ($\tau$; assuming $\tau \propto \nu^{-4}$ and referenced to $600$~MHz) in ms, spectral index ($\eta$), and spectral running ($r$, which allows for complex frequency structure to explain the limited bandwidth of some FRBs). In the case of a complex, multi-component burst, {\tt fitburst} measures individual TOA, $w$, $r$, and $\eta$ for each component, assuming scattering is constant among them. We use the spectral running and spectral index to deduce the central frequency ($\nu_\mathrm{c}$ in MHz) and bandwidth ($\Delta\nu$ in MHz) of each burst following Equation 1 by \cite{2021ApJS..257...59C}. Since scatter broadening and downward drifting can be difficult to distinguish, we determine whether the {\tt fitburst} fit with or without scatter broadening provides a more accurate solution for each burst using a reduced $\chi^2$ test and also by visually inspecting the fit residuals. For some complex bursts, we evaluate both scattering and non-scattering models on only the brightest burst component to better determine if scatter broadening is present. In cases where the fit without scattering represents the data better, we consider the width of the narrowest component to be an upper limit for the scattering timescale \citep{2025ApJ...979..160S, 2025ApJ...992..206C}.

Next, we calibrate the amplitude of our baseband data using daily observations of gain calibrators and correcting for the primary beam response of the telescope towards the FRB position \citep{Michilli2024}. After calibration, we compute the fluence, ${\cal F}_\nu$, by integrating over the 400$-$800~MHz frequency range. The fluence is then used to calculate the isotropic-equivalent spectral energy for each burst following \cite{2018MNRAS.480.4211M}:
\begin{equation}
E_\nu = \frac{4 \pi D_L^2}{(1+z)^{2+\rho}}{\cal F}_\nu\,, \label{eq:spec_energy}
\end{equation}
where $D_L$ is the luminosity distance to the FRB source, which we compute using the redshift $z=0.1839 \pm 0.0001$ \citep{2026Sci...391..280L} assuming the cosmological parameters from \citet{PlanckCollaboration_2020_A&A}, and $\rho$ is the FRB spectral index, which we set as $\rho=0$, consistent with our band-averaged fluence calculation.

We derive the polarization properties by applying the CHIME/FRB polarization pipeline \citep{2021ApJ...920..138M, 2024ApJ...968...50P}. We use RM-synthesis \citep{2005A&A...441.1217B} to measure an RM, requiring the peak polarized intensity in Faraday depth space to exceed $6\sigma$ \citep[after applying the {\tt RM-CLEAN} framework;][]{2009A&A...503..409H} to warrant a robust detection. For the bursts that do not meet this criterion, we provide only an upper limit on their linear polarization fraction ($L/I$) of $6/(\mathrm{S/N}(I))$, where $\mathrm{S/N}(I)$ is the signal-to-noise of the Stokes $I$ signal \citep{2024ApJ...968...50P}, and we do not measure an RM. For the bursts with an RM detection, we derotate the Stokes dynamic spectra using the measured RM and correct for potential instrumental polarization \citep[for details on this process, see][]{2021ApJ...920..138M, 2024ApJ...968...50P}; we then estimate the $L/I$ of these polarized bursts from the derotated Stokes Q and U spectra as $L/I = \sqrt{Q^2 + U^2}/I$.

\subsection{HyperFlash} \label{sec:obs_hyperflash}
We observed FRB~20220529A as part of the HyperFlash project (PI: O.~S.~Ould-Boukattine), designed to perform ultra high-cadence monitoring of repeating FRB sources. In our campaign targeting FRB~20220529A, two HyperFlash telescopes participated: the Toru\'n telescope (in Poland) and the RT-1 Westerbork telescope (in the Netherlands). In total, we observed for $678$~hours between 2023 June 06 and 2025 December 09. Toru\'n observed for a total of $14.98$~hours, with an observing band of $1380$–$1508$~MHz, achieving a detection threshold of $5.5$~Jy~ms. Westerbork observed with two observing bands: $300$–$356$~MHz for $55.98$~hours (detection threshold of $46.5$~Jy~ms) and $1207$–$1335$~MHz for $612.99$~hours (detection threshold of $6.6$~Jy~ms). We detected one FRB~20220529A burst with Westerbork at $1.4$~GHz on 2025 September 5 with a fluence of $10.9 \pm 2.2$~Jy~ms. Detailed background on the FRB search methodology and detection thresholds for the HyperFlash project can be found in \cite{2025arXiv250916374O}. 

\subsection{VLA} \label{sec:obs_vla}
The VLA is a radio telescope operated by the National Radio Astronomy Observatory in San Agustin, New Mexico. The telescope is comprised of 27 parabolic, 25-meter dishes that are arranged into three elongated arms with 9 dishes each. The VLA has 4 standard configuration sizes: A, B, C, and D, which have maximum baselines of 36.4, 11.4, 3.4, and 1.0~km, respectively. We conduct a search for continuum radio emission at the FRB~20220529A position in archival VLA observations (project code: 23A-385, PI: Ye Li) which targeted the source for 7~hours over six days in February 2023. These observations were conducted at L-band ($1-2$~GHz) while the VLA was in B-array configuration; thus, the synthesized beam size for the observations was $\sim 4.3''$.

\section{Results} \label{sec:results}
15 of the 16 bursts for which we have baseband data contain multiple, resolved components that often occupy distinct frequency ranges. In general, we do not visually see variation in propagation effects between components; thus we fit a single value of DM, RM, and $\tau$ for all components in a given burst. The spectral energy, $E_\nu$, is also computed across the full burst envelope of each FRB. Table \ref{tb:burst_props} summarizes the measured baseband properties of each burst (detection S/N, $E_\nu$, DM, RM, and $\tau$) and, when appropriate, of each subcomponent (TOA, $L/I$, $\nu_\mathrm{c}$, $\Delta\nu$, and $w$). 

\setlength{\tabcolsep}{1.3pt}
\setlength{\LTcapwidth}{1.0\textwidth}
\renewcommand{\arraystretch}{1}
\begin{center}
\begin{longtable}{cccccccccccc}
\caption{Summary of burst properties for 16 FRB 20220529A bursts detected by CHIME/FRB with baseband data.} \label{tb:burst_props}\\
\hline
\hline
Burst/ & S/N$^2$ & $E_\nu$ & DM & RM & $\tau$ at $600$~MHz & TOA$^3$ & $L/I$ & $\nu_\mathrm{c}$ & $\Delta\nu$ & $w$\\
Component$^1$ &  & (erg~Hz$^{-1}$) & (pc~cm$^{-3}$) & (rad~m$^{-2}$) & (ms) & (MJD) &  & (MHz) & (MHz) & (ms)\\
\hline
\endfirsthead

\multicolumn{7}{c}
{{\bfseries \tablename\ \thetable{} -- continued from previous page}} \\
\hline
Burst/ & S/N$^2$ & $E_\nu$ & DM & RM & $\tau$ at $600$~MHz & TOA$^3$ & $L/I$ & $\nu_\mathrm{c}$ & $\Delta\nu$ & $w$\\
Component$^1$ &  & (erg~Hz$^{-1}$) & (pc~cm$^{-3}$) & (rad~m$^{-2}$) & (ms) & (MJD) &  & (MHz) & (MHz) & (ms)\\
\hline
\endhead

\hline \multicolumn{2}{c}{{Continued on next page}} \\ \hline
\endfoot
\hline
\endlastfoot

Burst 1 & 16.77 & $9(1) \times 10^{30}$ & 246.40(8) & $-$ & $< 0.93(3)$ &  &  &  &  & \\
a &  &  &  &  &  & 59728.701470002 & $< 0.37(9)$ & 503 & 81 & 1.2(2)\\
b &  &  &  &  &  & 59728.701470131 & $< 0.21(3)$ & 450 & 79 & 0.93(3)\\
Burst 2 & 22.87 & $1.0(1) \times 10^{31}$ & 246.454(5) & $+43(1)$ & 1.43(9) &  &  &  &  & \\
a &  &  &  &  &  & 59745.653323534 & $< 0.24(4)$ & 712 & 115 & 0.852(5)\\
b &  &  &  &  &  & 59745.653323572 & $0.55(4)$ & 679 & 89 & 0.270(5)\\
c &  &  &  &  &  & 59745.653323687 & $< 0.25(7)$ & 467 & 134 & 1.7(2)\\
Burst 3$^4$ & 14.18 & $8.1(9) \times 10^{30}$ & 246.16(7) & $-$ & 0.37(1) &  &  &  &  & \\
a &  &  &  &  &  & 59771.581796623 & $< 0.41(9)$ & 705 & 64 & 0.489(1)\\
b &  &  &  &  &  & 59771.581796754 & $< 0.23(5)$ & 617 & 74 & 0.349(2)\\
c &  &  &  &  &  & 59771.581796811 & $< 0.5(2)$ & 425 & 54 & 0.333(5)\\
Burst 4 & 14.51 & $8(1) \times 10^{30}$ & 246.30(7) & $-$ & $< 0.7(1)$ &  &  &  &  & \\
a &  &  &  &  &  & 59774.572726902 & $< 0.5(2)$ & 717 & 103 & 2.3(3)\\
b$^5$ &  &  &  &  &  & 59774.572727107 & $< 1.4(4)$ & $> 800$ & $> 177$ & 2.5(4)\\
c &  &  &  &  &  & 59774.572727349 & $< 0.2(3)$ & 746 & 120 & 0.7(1)\\
d &  &  &  &  &  & 59774.572727554 & $< 0.22(3)$ & 679 & 89 & 2.7(1)\\
Burst 5 & 21.20 & $4.5(5) \times 10^{31}$ & 245.80(1) & $-4.4(4)$ & 0.41(2) &  &  &  &  & \\
a$^5$ &  &  &  &  &  & 59923.171025075 & $< 0.8(2)$ & $> 800$ & $> 302$ & 1.851(1)\\
b &  &  &  &  &  & 59923.171025239 & $< 0.5(1)$ & 616 & 175 & 0.164(1)\\
c &  &  &  &  &  & 59923.171025263 & $< 1.0(4)$ & 590 & 189 & 0.923(1)\\
d &  &  &  &  &  & 59923.171025328 & $0.48(3)$ & 501 & 106 & 0.846(1)\\
Burst 6 & 15.83 & $3.8(4) \times 10^{31}$ & 245.807(5) & $-$ & $< 0.45(3)$ &  &  &  &  & \\
a &  &  &  &  &  & 59927.153918110 & $<0.6(2)$ & 681 & 59 & 0.45(3)\\
b &  &  &  &  &  & 59927.153918177 & $< 0.6(3)$ & 649 & 115 & 0.58(2)\\
c &  &  &  &  &  & 59927.153918304 & $< 0.22(9)$ & 587 & 79 & 0.66(1)\\
d &  &  &  &  &  & 59927.153918348 & $< 0.42(8)$ & 566 & 88 & 0.58(2)\\
e &  &  &  &  &  & 59927.153918372 & $< 0.17(3)$ & 561 & 85 & 0.84(3)\\
Burst 7 & 18.48 & $5.3(7) \times 10^{30}$ & 245.83(6) & $+46.7(6)$ & $<0.466(5)$ & 59929.150721482 & $0.37(3)$ & 605 & 123 & 0.466(5)\\
Burst 8$^4$ & 14.53 & $2.5(3) \times 10^{31}$ & 245.494(7) & $-$ & $< 0.54(1)$ &  &  &  &  & \\
a &  &  &  &  &  & 60100.682953496 & $< 0.29(7)$ & 614 & 87 & 0.67(3)\\
b &  &  &  &  &  & 60100.682953531 & $< 0.30(6)$ & 620 & 85 & 0.54(1)\\
c$^6$ &  &  &  &  &  & 60100.682953727 & $< 0.26(5)$ & 572 & 80 & 0.71(1)\\
d$^6$ &  &  &  &  &  & 60100.682953785 & $< 0.37(3)$ & 518 & 94 & 0.79(2)\\
Burst 9 & 12.21 & $4.5(6) \times 10^{30}$ & 245.45(1) & $-$ & $< 0.6(1)$ &  &  &  &  & \\
a &  &  &  &  &  & 60114.642741616 & $< 0.3(2)$ & 691 & 123 & 0.6(1)\\
b &  &  &  &  &  & 60114.642741666 & $< 0.8(3)$ & 618 & 84 & 1.2(4)\\
c &  &  &  &  &  & 60114.642741802 & $< 0.4(1)$ & 513 & 111 & 2.3(1)\\
Burst 10 & 23.76 & $2.2(2) \times 10^{31}$ & 244.86(5) & $-0.4(5)$ & $< 0.583(9)$ &  &  &  &  & \\
a$^6$ &  &  &  &  &  & 60232.318466390 & $0.34(1)$ & 631 & 96 & 0.695(9)\\
b &  &  &  &  &  & 60232.318466421 & $< 0.30(6)$ & 604 & 96 & 0.592(2)\\
c &  &  &  &  &  & 60232.318466589 & $< 0.17(3)$ & 436 & 73 & 0.583(9)\\
Burst 11 & 40.29 & $4.4(4) \times 10^{31}$ & 244.872(9) & $-2.2(4)$ & $< 0.586(4)$ &  &  &  &  & \\
a &  &  &  &  &  & 60236.312472855 & $< 0.32(2)$ & 610 & 73 & 0.586(4)\\
b$^6$ &  &  &  &  &  & 60236.312473007 & $0.24(1)$ & 444 & 72 & 1.066(8)\\
Burst 12 & 18.65 & $8(1) \times 10^{30}$ & 244.91(2) & $-$ & $< 0.90(3)$ &  &  &  &  & \\
a &  &  &  &  &  & 60237.306339305 & $< 0.3(1)$ & 618 & 95 & 0.9(1)\\
b &  &  &  &  &  & 60237.306339502 & $< 0.22(6)$ & 449 & 87 & 0.90(3)\\
c &  &  &  &  &  & 60237.306339658 & $< 0.24(6)$ & 431 & 75 & 1.22(6)\\
Burst 13$^7$ & 22.10 & $3.6(5) \times 10^{31}$ & 244.82(4) & $-3.5(5)$ & 0.43(1) &  & $0.44(4)$ &  &  & \\
a &  &  &  &  &  & 60243.286752607 & $-$ & 427 & 46 & 0.85(7)\\
b &  &  &  &  &  & 60243.286752630 & $-$ & 424 & 47 & 0.25(4)\\
c &  &  &  &  &  & 60243.286752652 & $-$ & 420 & 30 & 0.16(9)\\
Burst 14 & 14.09 & $5.8(8) \times 10^{30}$ & 245.67(4) & $-$ & $< 0.87(1)$ &  &  &  &  & \\
a &  &  &  &  &  & 60279.192819393 & $< 0.20(4)$ & 650 & 91 & 0.87(1)\\
b &  &  &  &  &  & 60279.192819430 & $< 0.3(1)$ & 587 & 84 & 1.06(3)\\
Burst 15 & 12.09 & $1.0(1) \times 10^{31}$ & 244.3(2) & $-$ & $< 1.47(2)$ &  &  &  &  & \\
a &  &  &  &  &  & 60591.340086683 & $< 0.26(4)$ & 580 & 104 & 1.47(2)\\
b$^5$ &  &  &  &  &  & 60591.340086752 & $< 0.3(1)$ & $< 400$ & $> 133$ & 2.14(6)\\
Burst 16 & 20.61 & $2.7(3) \times 10^{31}$ & 243.7(1) & $+7(1)$ & $1.96(1)^8$ &  &  &  &  & \\
a &  &  &  &  &  & 60898.497873852 & 0.28(1) & 690 & 60 & 1.17(1)\\
b &  &  &  &  &  & 60898.497873913 & $<0.20(3)$ & 623 & 110 & 0.55(2)\\
\hline
\hline
\multicolumn{11}{l}{All uncertainties are quoted at the 1-$\sigma$ level.} \\
\multicolumn{11}{l}{$^{1}$Burst components appear in chronological order. Properties for single component bursts are presented in a single row.} \\
\multicolumn{11}{l}{$^{2}$We report the S/N measured by the real-time CHIME/FRB search \citep{2018ApJ...863...48C}.} \\
\multicolumn{11}{l}{$^{3}$All TOAs are topocentric at CHIME, and are referenced to 400~MHz after dedispersing to the DM shown in column 4.} \\
\multicolumn{11}{l}{$^{4}$Some components are too faint to reliably constrain their burst properties; they are not included in the table.} \\
\multicolumn{11}{l}{$^{5}$For burst components with a best-fit central frequency that lies outside of $400-800$~MHz, we report an upper/lower limit} \\
\multicolumn{11}{l}{on $\nu_c$ and a lower limit on $\Delta \nu$ corresponding to the bandwidth occupied by the component within the CHIME/FRB band.} \\
\multicolumn{11}{l}{$^6$Likely comprised of narrower, overlapping components that are not cleanly distinguishable. Burst properties of the} \\
\multicolumn{11}{l}{broader emission envelope are provided.} \\
\multicolumn{11}{l}{$^7$Polarization properties are computed across the full burst envelope to increase linearly polarized S/N and obtain an RM.} \\
\multicolumn{11}{l}{$^8$We consider this scattering timescale a tentative measurement as we cannot definitely rule out downward drifting} \\
\multicolumn{11}{l}{morphology that partially overlaps with (masked) radio frequency interference at $\sim 620-640$~MHz.} \\
\end{longtable}
\end{center}

\begin{figure*}[ht!]
\begin{center}
    \includegraphics[width=0.235\textwidth]{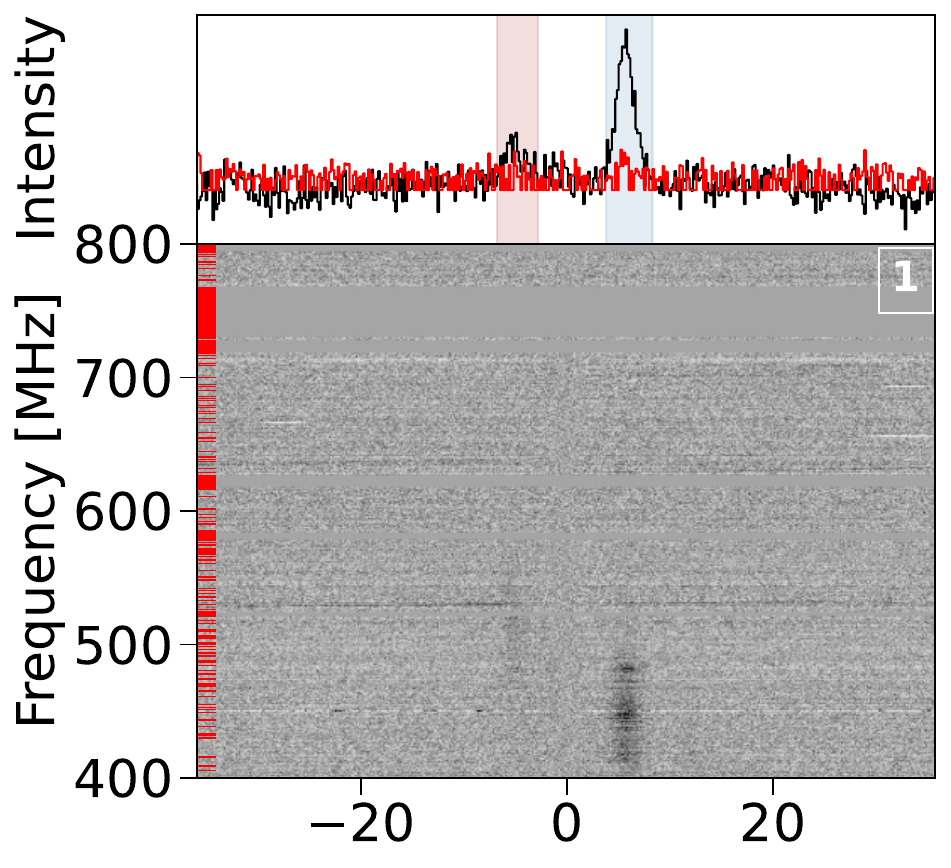}
    \includegraphics[width=0.224\textwidth]{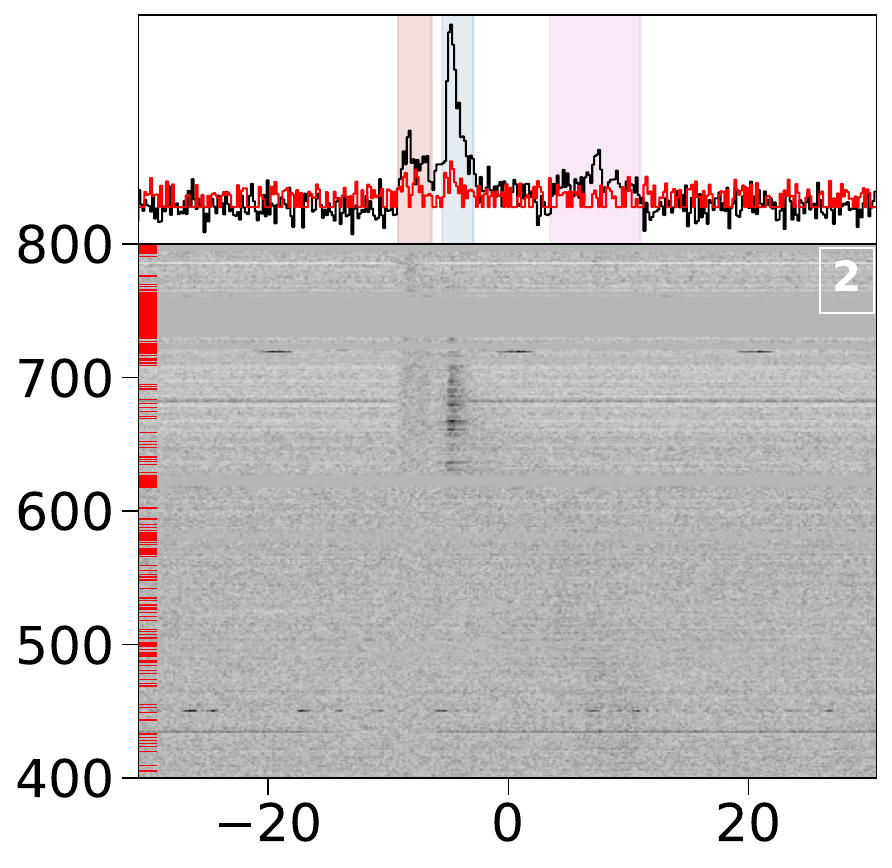}
    \includegraphics[width=0.224\textwidth]{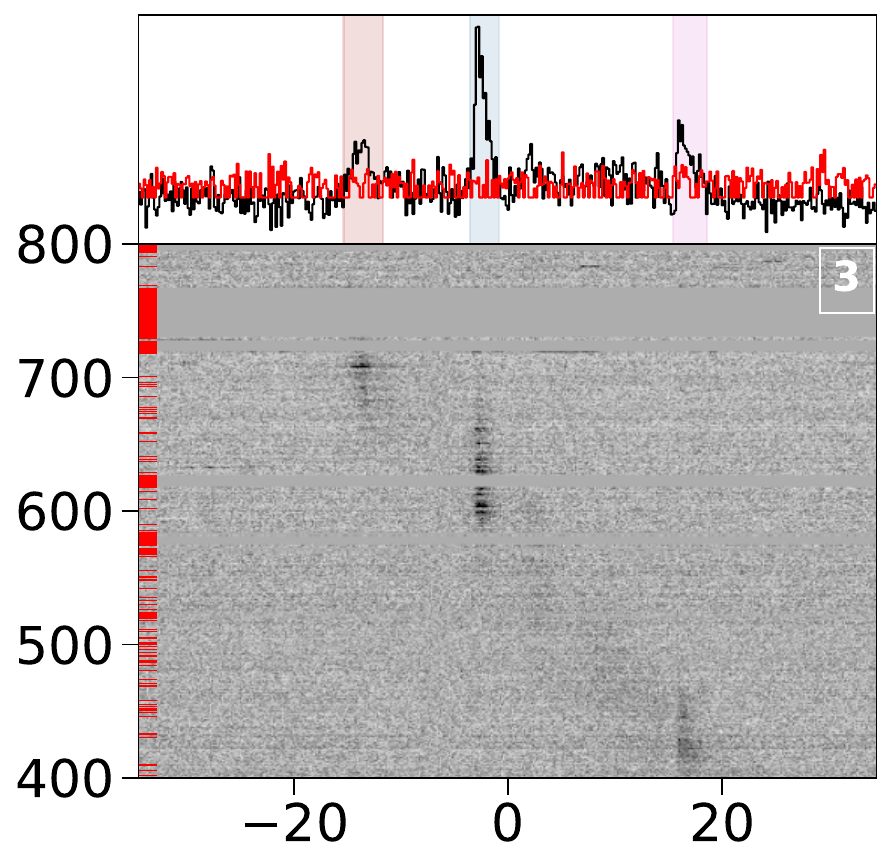}
    \includegraphics[width=0.224\textwidth]{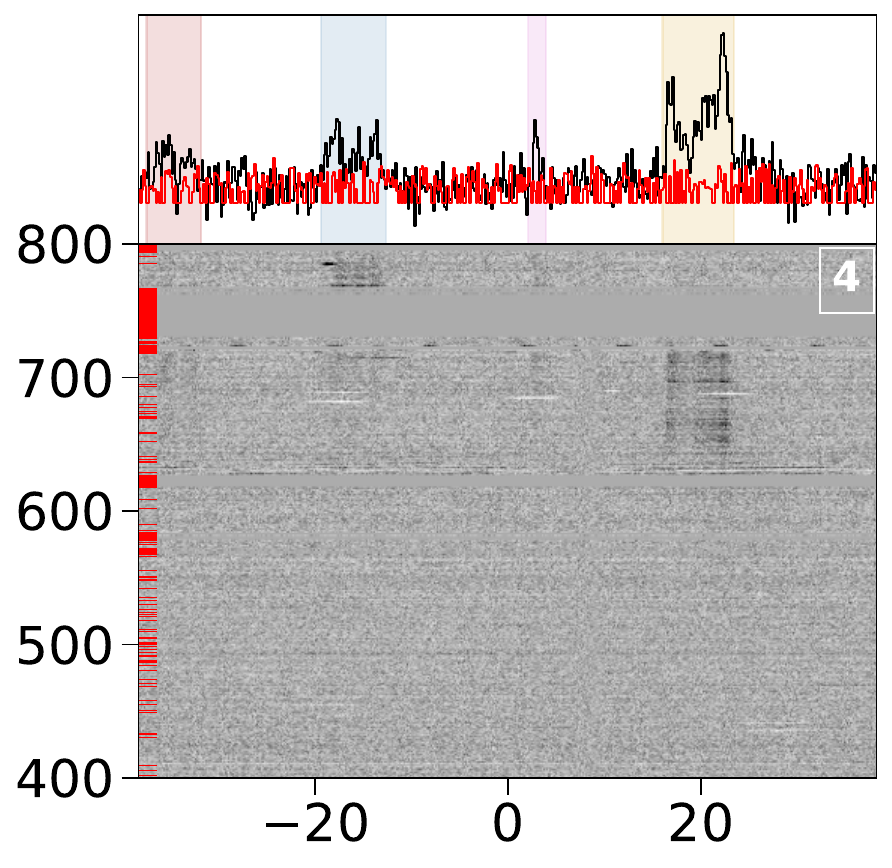}

    \includegraphics[width=0.235\textwidth]{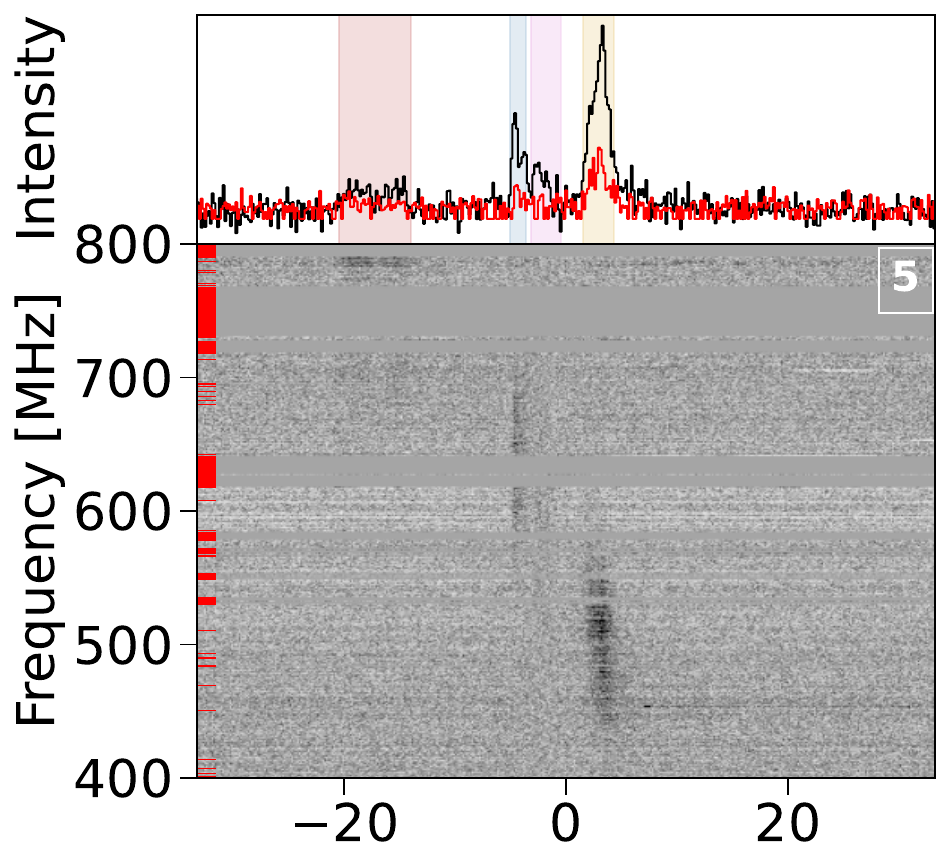}
    \includegraphics[width=0.224\textwidth]{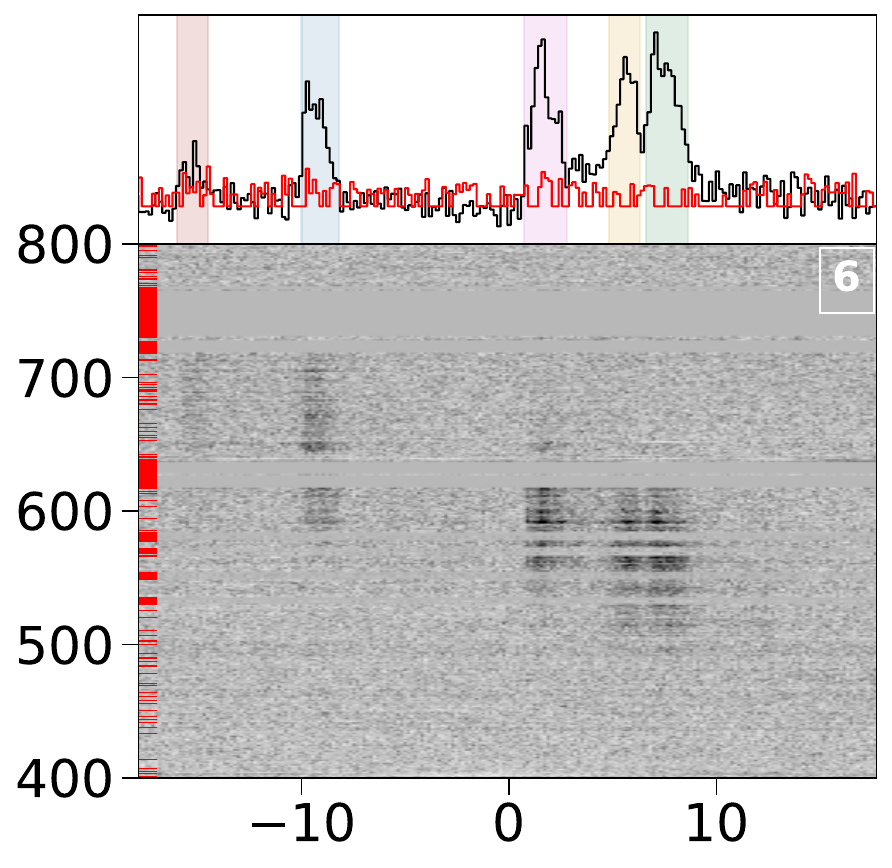}
    \includegraphics[width=0.224\textwidth]{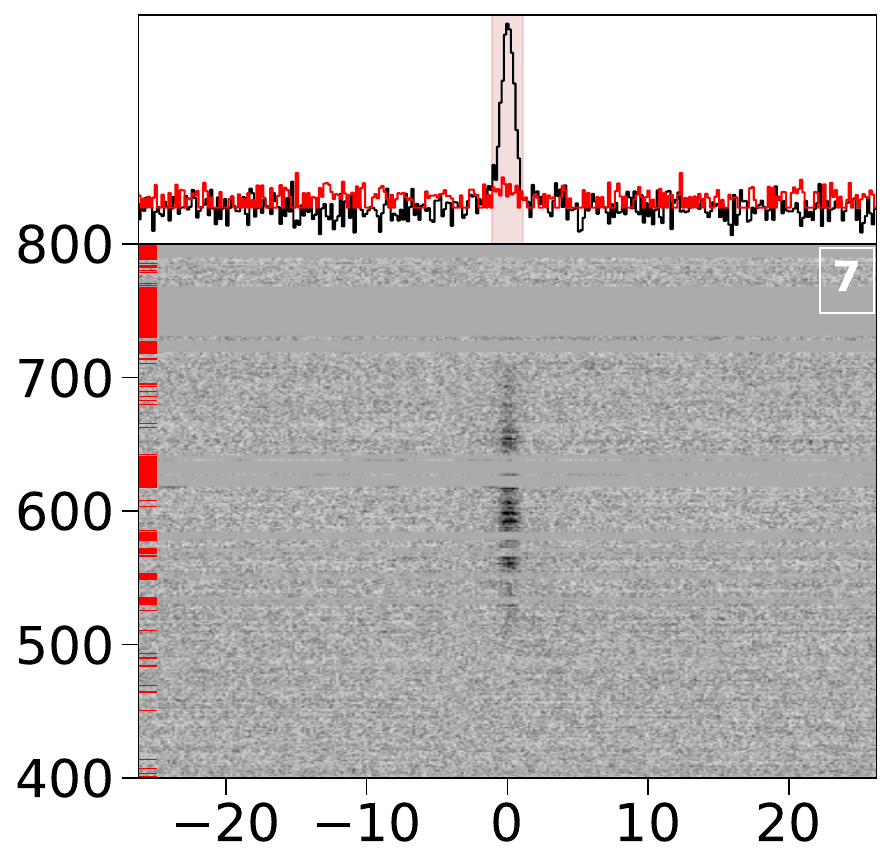}
    \includegraphics[width=0.224\textwidth]{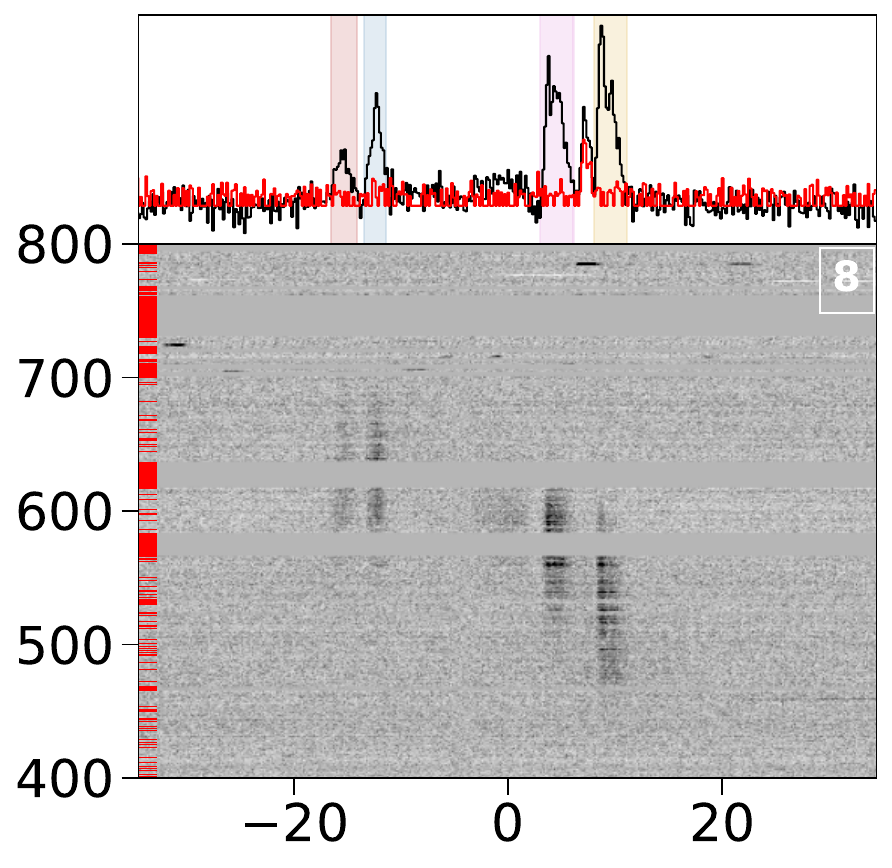}
    
    \includegraphics[width=0.235\textwidth]{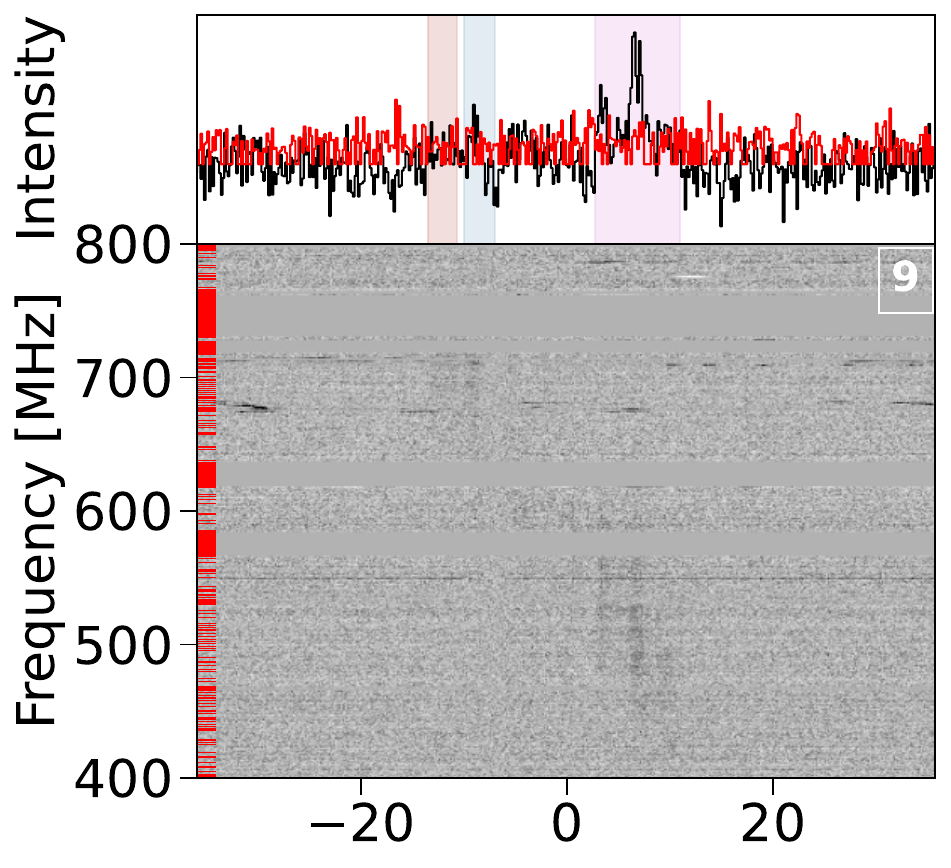}
    \includegraphics[width=0.224\textwidth]{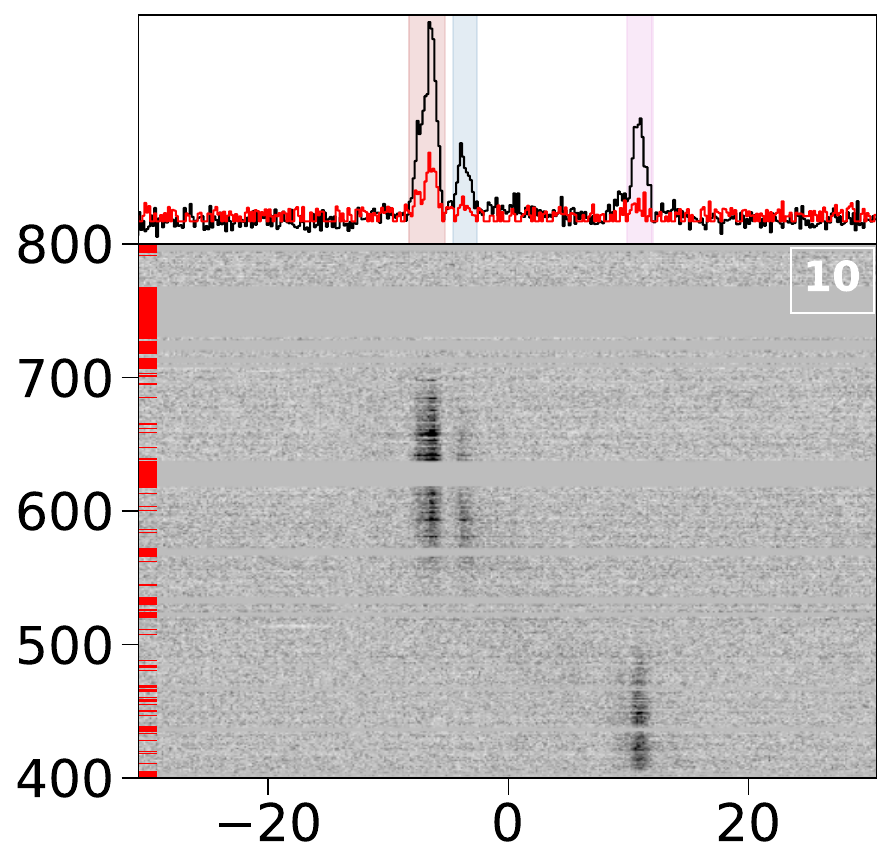}
    \includegraphics[width=0.224\textwidth]{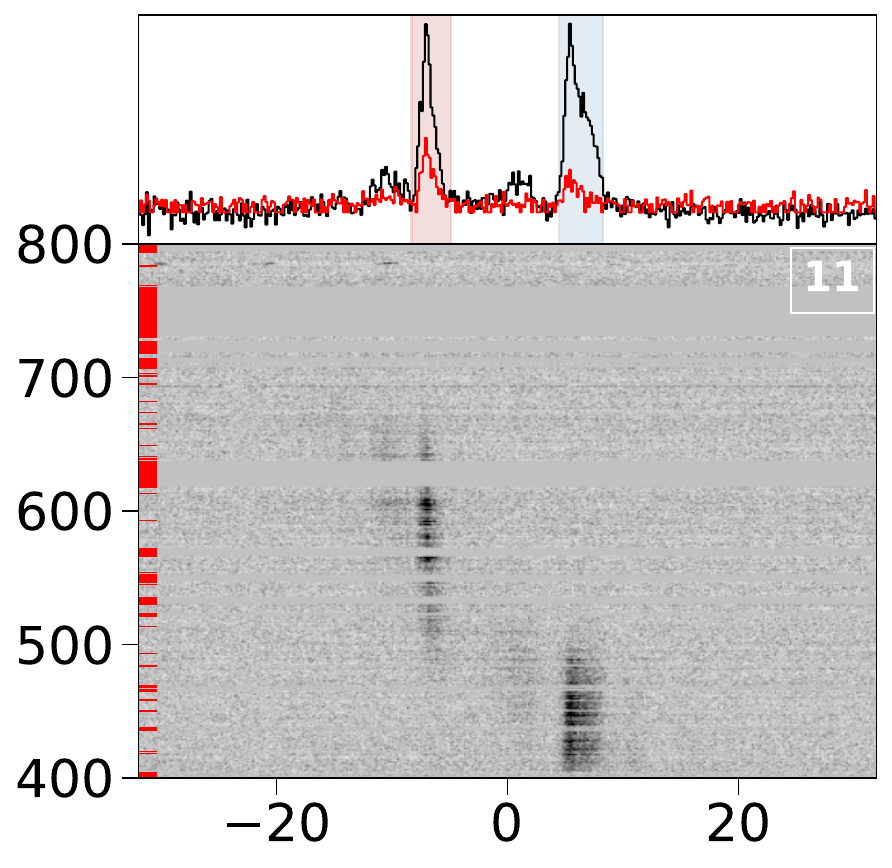}
    \includegraphics[width=0.224\textwidth]{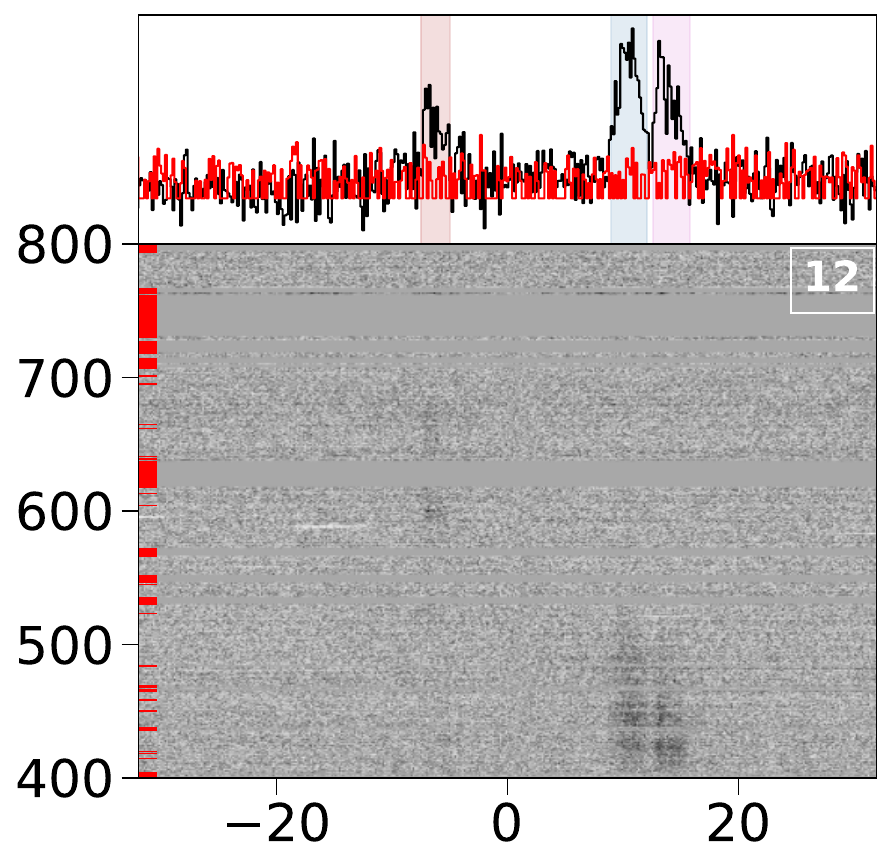}
    
    \includegraphics[width=0.235\textwidth]{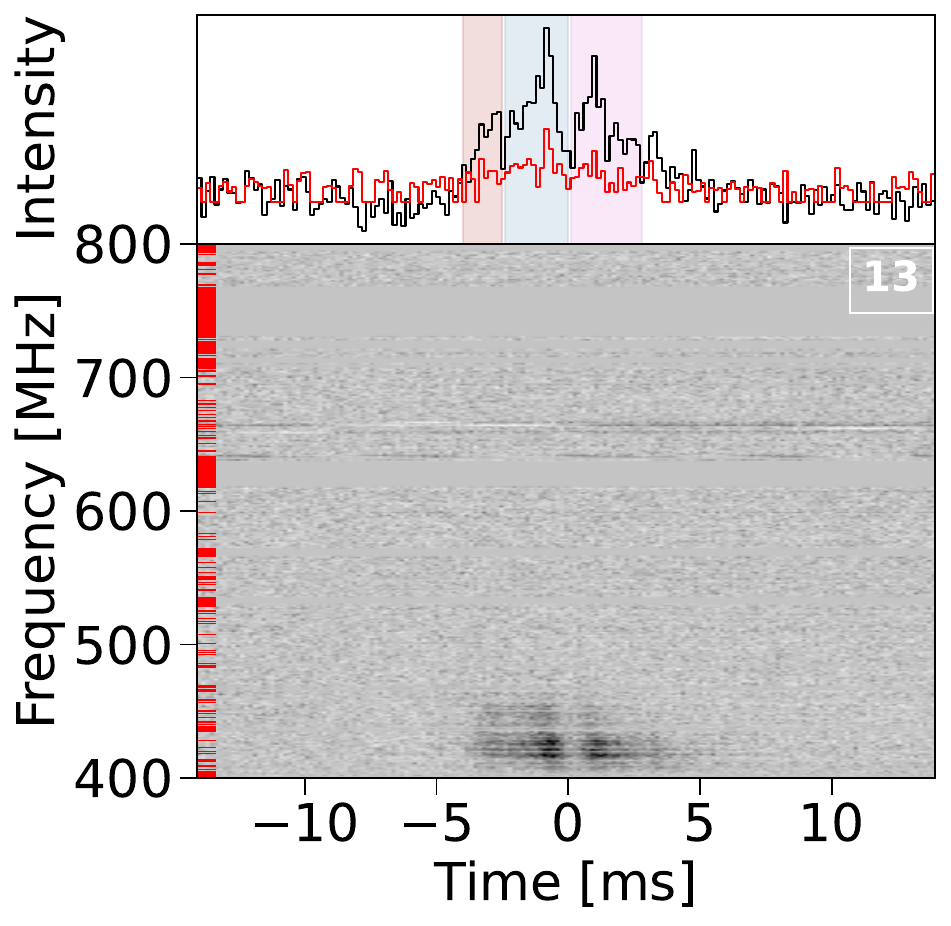}
    \includegraphics[width=0.224\textwidth]{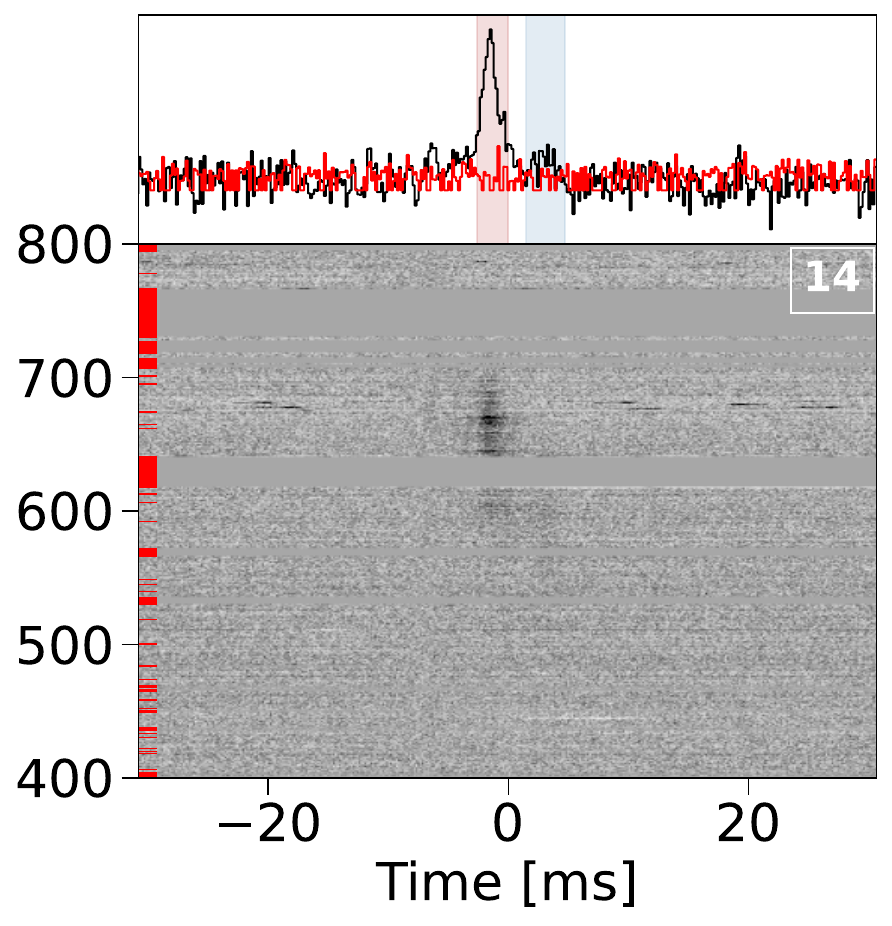}
    \includegraphics[width=0.224\textwidth]{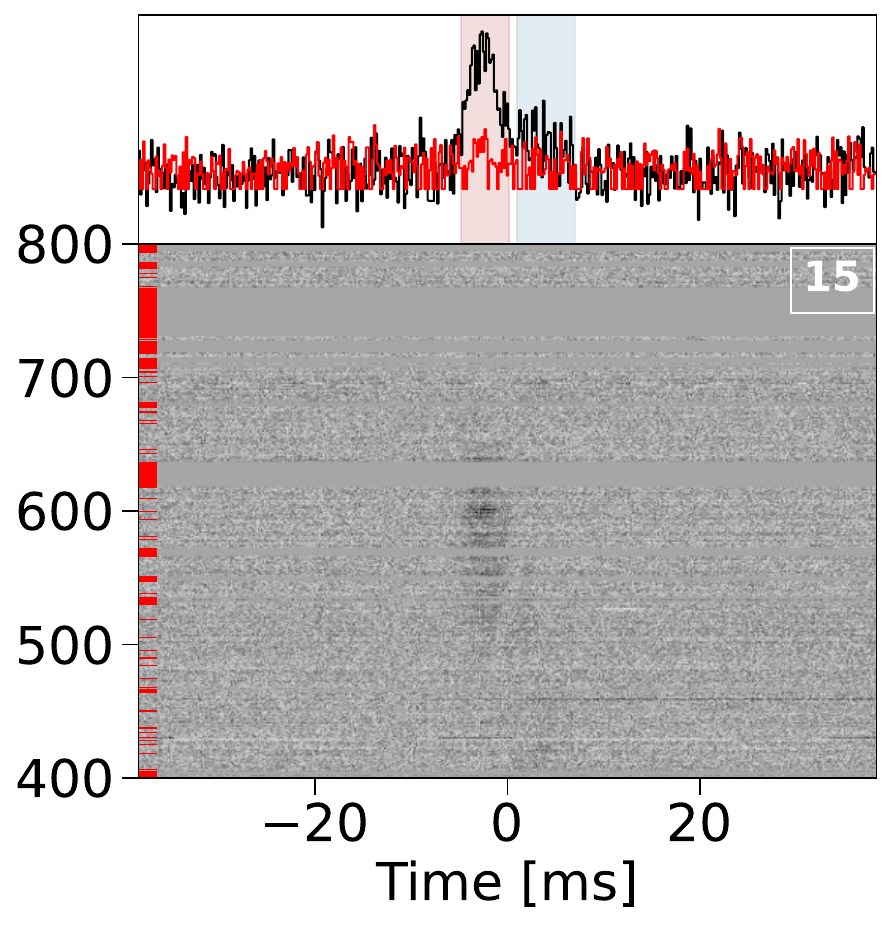}
    \includegraphics[width=0.224\textwidth]{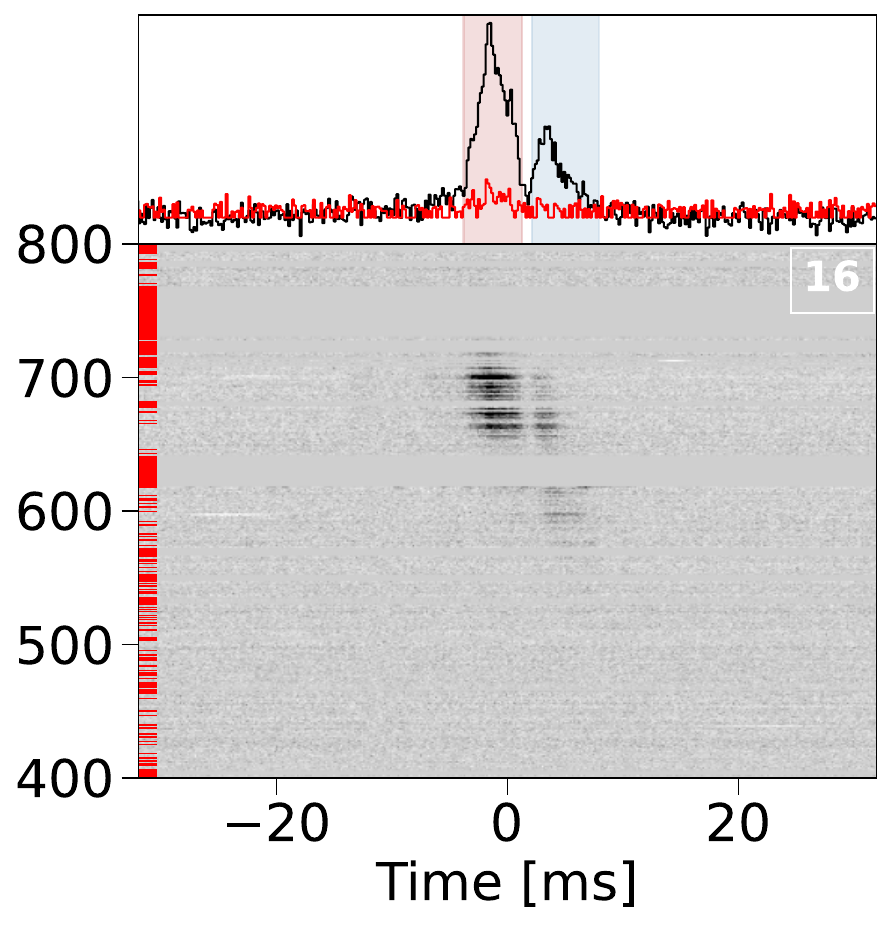}
    \caption{Total intensity dynamic spectra, dedispersed to their best-fit DMs, for the 16 FRB~20220529A bursts with baseband data detected by CHIME/FRB. All dynamic spectra have a frequency resolution of $390.625$~kHz and have been downsampled to a common time resolution of $163.84~\mu\mathrm{s}$ to ensure sufficient S/N in fainter bursts. Frequency-averaged total intensity (black) and linearly polarized intensity (red) profiles are plotted above the dynamic spectra. Individual burst components are highlighted by shaded regions (in chronological order for multiple components: red, blue, purple, yellow, and green) overlaid on the intensity profiles. Channels masked out due to radio frequency interference are highlighted in red on the left of the dynamic spectra.}
    \label{fig:waterfalls}
\end{center}
\end{figure*}

\subsection{Dispersion} \label{sec:results_dm}
\begin{figure*}
    \centering
    \includegraphics[width=0.93\textwidth]{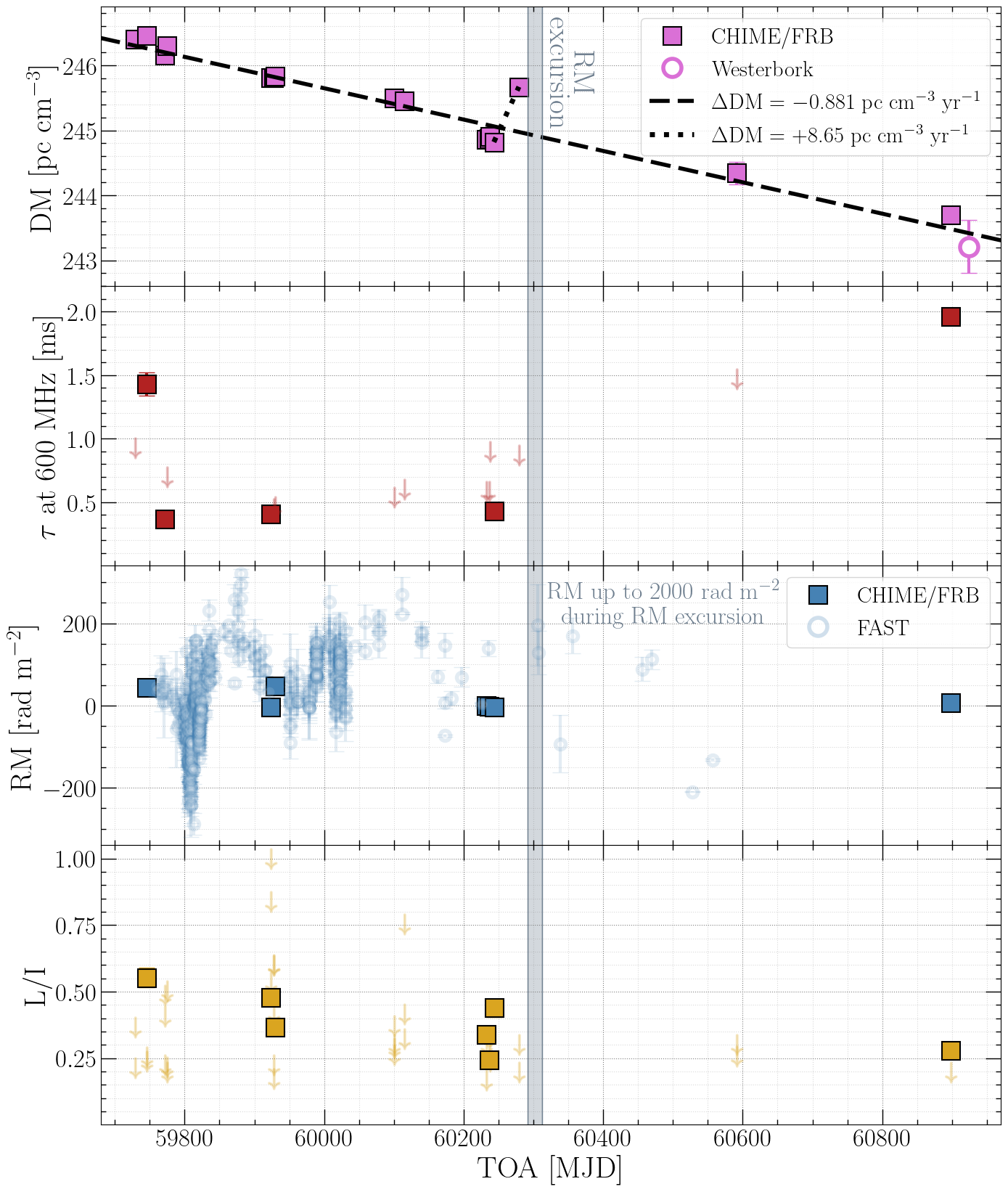}
    \caption{Temporal evolution of the DM, $\tau$, RM, and $L/I$ in FRB~20220529A as seen by CHIME/FRB (squares). Uncertainties in these parameters are smaller than the markers. The ``RM excursion'' \citep{2026Sci...391..280L} is shaded in gray. In the top panel, the best-fit linear DM decline is plotted as a dashed line and the slope of the rise in DM preceding the RM excursion is depicted with a dotted line. The DM of a burst detected by the RT-1 Westerbork radio telescope is plotted as an open purple circle. In the second and the bottom panels, upper limits on $\tau$ and $L/I$, respectively, are plotted as downward arrows. To show how the CHIME/FRB RMs fit into the broader RM evolution of FRB~20220529A, we overplot the FAST RM measurements by \citep{2026Sci...391..280L} as blue open circles in the third panel; the RM excursion peak falls outside the y-axis range in this panel.} 
    \label{fig:time_var}
\end{figure*}

We determine the structure-maximizing DM for all FRB~20220529A bursts in our sample (see Table \ref{tb:burst_props}), and we plot their total intensity dynamic spectra dedispersed to their respective DMs in Figure \ref{fig:waterfalls}. In the top panel of Figure \ref{fig:time_var}, we plot the temporal evolution of the DM. Two features are immediately apparent: (i) a sustained decrease in the DM over $\sim 3.2$~years and (ii) a abrupt rise in the DM immediately preceding the ``RM excursion'' of $\sim 2000~\mathrm{rad}~\mathrm{m}^{-2}$ reported by \cite{2026Sci...391..280L}, which spans $60292 < \mathrm{MJD} < 60312$ and is shown as a shaded gray region in Figure \ref{fig:time_var}.

We find that the DM decreases from $246.40\pm0.08~\mathrm{pc}~\mathrm{cm}^{-3}$ to $243.7\pm0.1~\mathrm{pc}~\mathrm{cm}^{-3}$ over the course of $1170$ days. A simple linear fit, done using the {\tt scipy.optimize} package, reveals a rate of change in the observer-frame DM of $\mathrm{dDM}/\mathrm{d}t = -0.881\pm0.001~\mathrm{pc}~\mathrm{cm}^{-3}~\mathrm{year}^{-1}$ (overplotted as a dashed black line in the top panel of Figure \ref{fig:time_var}). All but one of the bursts closely agree with this DM evolution with only small variations about the best-fit line. While the model fit uncertainty on the slope is extremely small ($\sim 10^{-5}~\mathrm{pc}~\mathrm{cm}^{-3}~\mathrm{year}^{-1}$), we adopt a more conservative $\pm0.001~\mathrm{pc}~\mathrm{cm}^{-3}~\mathrm{year}^{-1}$ uncertainty that encompasses our average DM residuals ($\sim 0.1~\mathrm{pc}~\mathrm{cm}^{-3}$) about the best-fit line. In the rest frame, the rate of change in the DM is $-1.235\pm0.001~\mathrm{pc}~\mathrm{cm}^{-3}~\mathrm{year}^{-1}$. For the single burst detected from FRB~20220529A detected by Westerbork RT-1, we measure a DM of $243.2 \pm 0.4~\mathrm{pc}~\mathrm{cm}^{-3}$ (for details see Appendix~\ref{sec:appendix_B}), consistent with the decreasing DM trend observed in the CHIME/FRB bursts (Figure~\ref{fig:time_var}). 

The higher frequency and lower time resolution FAST data are more prone to degeneracies between DM and burst morphology, leading to perceived intraday DM fluctuations of $\sim1-10~\mathrm{pc}~\mathrm{cm}^{-3}$ that are unlikely to be explained by electron column density variations \citep{2026Sci...391..280L}. This serves as an unmodeled systematic error on the FAST-measured DMs, making it impossible to detect gradual variations of $\sim1~\mathrm{pc}~\mathrm{cm}^{-3}~\mathrm{year}^{-1}$ over only a few-year baseline. To test the robustness of our measured DM decline, we perform Monte Carlo simulations in which 16 DMs are drawn from a Gaussian distribution matching the FAST DM measurements, assigned CHIME/FRB-like uncertainties ($\sim 0.1~\mathrm{pc}~\mathrm{cm}^{-3}$), and fit with a linear trend. Over $10^6$ trials, no simulated realization yields a more statistically significant linear trend (as measured by the Bayesian Information Criterion) than that observed in the CHIME/FRB data. This approach is developed in detail by Cook et al. ({\it in prep}).

Thirteen days before the reported RM excursion, the DM increased by $\Delta \mathrm{DM} = 0.85 \pm 0.04~\mathrm{pc}~\mathrm{cm}^{-3}$ between MJDs 60243 and 60279, corresponding to an extrapolated rate of change of $\mathrm{dDM}/\mathrm{d}t \geq +8.65~\mathrm{pc}~\mathrm{cm}^{-3}~\mathrm{year}^{-1}$. This point represents a $\sim 4.2\sigma$ deviation from the long-term linear DM decline. Owing to sparse sampling during this 36-day interval, we cannot constrain the detailed DM-time profile or peak DM. The next CHIME/FRB detection nearly a year later ($\mathrm{MJD}~60591$) is consistent with the observed long-term trend of $\mathrm{dDM}/\mathrm{d}t = -0.881\pm0.001~\mathrm{pc}~\mathrm{cm}^{-3}~\mathrm{year}^{-1}$. The RM increase to $1977\pm84~\mathrm{rad}~\mathrm{m}^{-2}$ was observed on MJD 60292 \citep{2026Sci...391..280L}; although the RM peak may have occurred earlier due to non-detections by FAST in the preceding $\sim57$ days. We conclude that the temporal proximity of the DM and RM changes suggests a common physical origin.

\subsection{Scattering} \label{sec:results_scat}
We measure scattering in 5 of the 16 CHIME/FRB bursts, and place upper limits for the remaining 11 (Table~\ref{tb:burst_props}). The scattering timescales, shown as a function of time in Figure \ref{fig:time_var} (second panel), vary by $\gtrsim 1$~ms on week-long timescales and potentially by $>2$~ms over a multi-year period.\footnote{See the table note related to the scattering timescale of burst 16 in Table \ref{tb:burst_props}.} Given that the predicted Milky Way contribution at 600~MHz is orders of magnitude smaller \citep[$\sim 10^{-3}$~ms in NE2001;][]{2002astro.ph..7156C,Ocker_2024_RNAAS}, the observed scattering is likely dominated by plasma local to the FRB source.

\subsection{Polarimetry} \label{sec:results_pol}
We measure a RM for 7 of the CHIME/FRB bursts (using their brightest components; Table~\ref{tb:burst_props}). The remaining burst components are not sufficiently linearly polarized for a robust RM detection (i.e., $< 6\sigma$) and are deemed ``unpolarized''. As at most one polarized component is detected per burst (usually the highest S/N component), we adopt its RM as representative of the burst. The RMs span $-4.4 \pm 0.4~\mathrm{rad}~\mathrm{m}^{-2}$ to $+46.7 \pm 0.6~\mathrm{rad}~\mathrm{m}^{-2}$ in the observer frame. Comparing to the more densely sampled FAST RM profile \citep{2026Sci...391..280L}, our RMs on MJDs 59745, 59923, and 59929 qualitatively follow the quasi-oscillatory long-term evolution. Notably, the RM increased by $51.1\pm0.7~\mathrm{rad}~\mathrm{m}^{-2}$ over just 7~days (MJDs~$59923–59929$), consistent with day-scale RM changes seen in FAST outside of the RM excursion. Between MJDs~$60232-60243$, RMs near $0~\mathrm{rad}~\mathrm{m}^{-2}$ are $\sim 140~\mathrm{rad}~\mathrm{m}^{-2}$ smaller than a FAST detection on MJD~$60236$, suggesting possible rapid fluctuations in the local magnetoionic plasma. We do not see a rapid RM rise corresponding to the RM excursion in \cite{2026Sci...391..280L}, implying it began after $\mathrm{MJD}~60243.28675$. We show the temporal evolution of the RM and $L/I$ in the bottom two panels of Figure \ref{fig:time_var}. We caution that the RM of burst 10 is consistent with $0~\mathrm{rad}~\mathrm{m}^{-2}$ and may, in part, reflect instrumental polarization.

The linear polarization fractions of the polarized bursts do not show any clear temporal evolution. However, FRB~20220529A consistently shows a low level of linear polarization at $400-800$~MHz, with polarized components ranging over $0.24\pm0.01 < L/I < 0.55\pm0.04$, and upper limits as low as $L/I < 0.17 \pm 0.03$. The $L/I$ of FRB~20220529A is significantly below the median $L/I$ for both non-repeating \citep[$L/I \sim 0.65$;][]{2024ApJ...968...50P} and repeating FRBs \citep[$L/I \sim 0.55$;][]{2025ApJ...982..154N}, respectively. The $L/I$ of FRB~20220529A bursts in the CHIME/FRB band are systematically lower than the $L/I$ observed at L-band by FAST, which have a median of $L/I \sim 0.8$ (see Figure \ref{fig:depol}; we discuss this depolarization in the following Section).

\subsubsection{Spectral Depolarization} \label{sec:results_depol}
As FRB emission passes through an inhomogeneous magnetoionic plasma, the signal becomes scattered to a finite angular size, undergoing differential Faraday rotation due to foreground spatial fluctuations in electron density and/or magnetic field. Averaging over these variations reduces the observed linear polarization with wavelength, $L/I (\lambda)$, relative to the intrinsic $L/I$ at a reference wavelength $\lambda=0$~m, $(L/I)_\mathrm{int}$:
\begin{equation}
L/I (\lambda) = (L/I)_\mathrm{int}~\mathrm{exp} \left(-2 \lambda^4 \sigma_\mathrm{RM}^2 \right)\,, \label{eq:beam_depol}
\end{equation}
where $\sigma_\mathrm{RM}$ is a measure of the inhomogeneity in the scattering medium \citep{1966MNRAS.133...67B}. While alternative spectral depolarization models exist \citep[e.g.,][]{1991MNRAS.250..726T, 2022MNRAS.510.4654B}, we adopt Equation \ref{eq:beam_depol} to facilitate direct comparison with previous FRB studies.

\begin{figure*}
    \centering
    \includegraphics[width=0.98\textwidth]{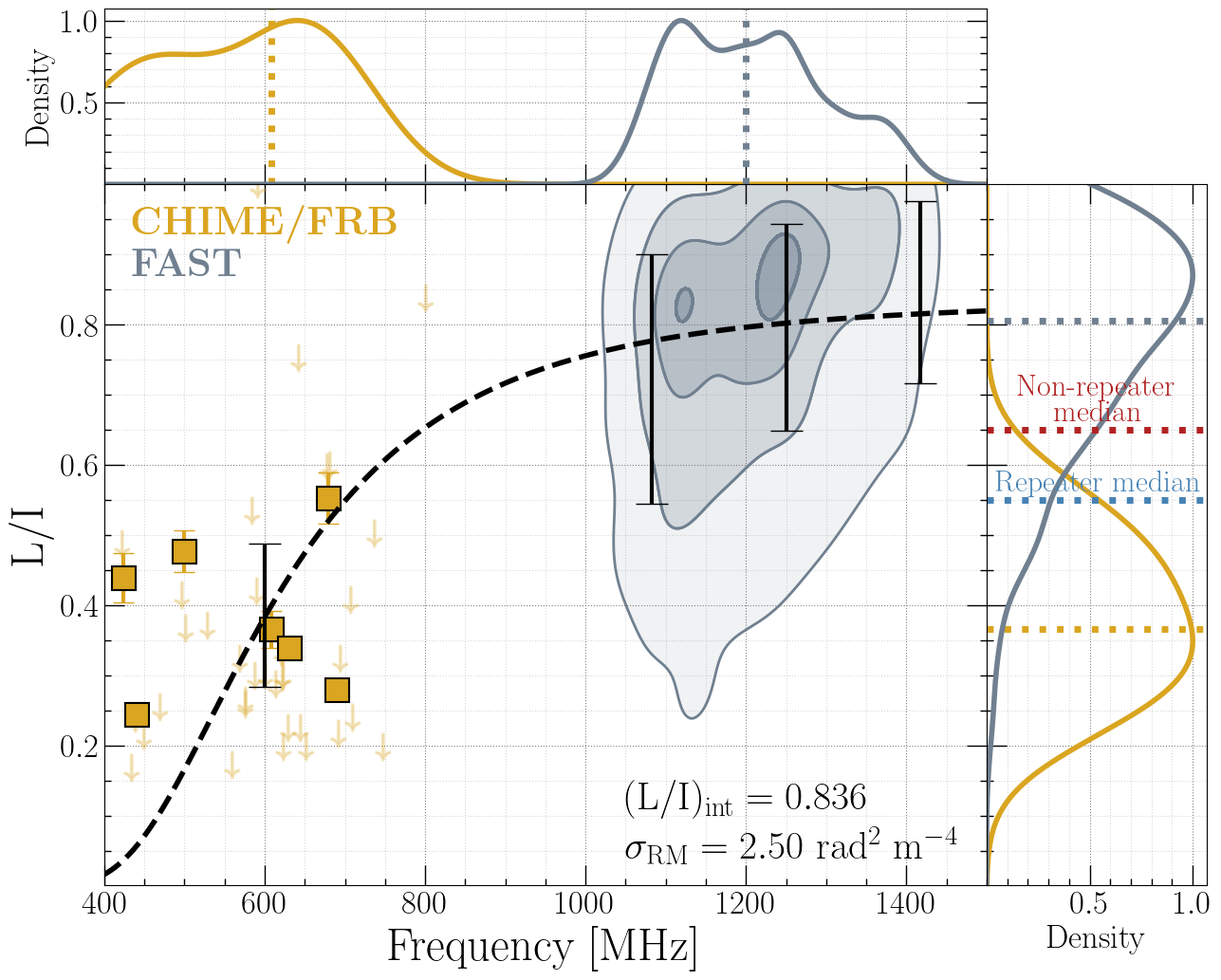}
    \caption{Linear polarization fraction versus central frequency of FRB~20220529A bursts reported by CHIME/FRB and FAST. Polarized bursts observed by CHIME/FRB are plotted as gold squares and $L/I$ upper limits are shown as downward arrows. The large quantity of FAST observations are depicted as a gray density map with increasing opacity at discrete contour levels of 5\%, 33\%, 67\%, and 95\%, for visual clarity. Bursts during the RM excursion have been removed from the FAST data. The best-fit model is overplotted as a black dashed line and the fit parameters are presented in the bottom right of the main panel. The binned $L/I$ data on which the depolarization model was fit are plotted as vertical black bars centered on the mean binned values and spanning their standard deviations. Upper limits on $L/I$ are not used in the depolarization model fitting. Smoothed kernel density estimates of the central frequency and $L/I$ distributions for polarized bursts observed by each instrument are plotted in gold and gray, respectively, above and to the right of the main panel. Gold and gray dotted lines in these panels represent the medians of the corresponding distributions. Median $L/I$ for polarized CHIME/FRB repeaters \citep{2025ApJ...982..154N} and non-repeaters \citep{2024ApJ...968...50P} are overplotted as blue and red dotted lines, respectively, in the right-most panel.} 
    \label{fig:depol}
\end{figure*}

We bin the $L/I$ measurements of FRB~20220529A from both CHIME/FRB and FAST in four frequency sub-bands ($400-800$~MHz, $1000-1166$~MHz, $1167-1332$~MHz, and $1333-1500$~MHz) and fit Equation~\ref{eq:beam_depol} using {\tt scipy.optimize}. Note we exclude FAST detections during the RM excursion episode which show significantly lower $L/I$. The best fit is shown in Figure~\ref{fig:depol} with depolarization parameters: $(L/I)_\mathrm{int} = 0.836 \pm 0.001$ and $\sigma_\mathrm{RM} = 2.50 \pm 0.02~\mathrm{rad}~\mathrm{m}^{-2}$. These statistical uncertainties underestimate the observed variance in $L/I$, which may reflect temporal changes in the depolarizing environment or intrinsic $L/I$ variability between bursts (see Section~\ref{sec:discussion_depol}).

In Appendix \ref{a:depol}, we derive the rest frame $\sigma_\mathrm{RM}$, $|\mathrm{RM}|$, and $\tau$ for all depolarizing FRBs in the literature \citep{2022Sci...375.1266F, 2025MNRAS.tmp.1894U}, including FRB~20220529A, and assess correlations among these properties. We find marginally significant positive correlations between $\sigma_\mathrm{RM}-|\mathrm{RM}|$ (p-value of 0.006: $\sim2.5\sigma$), and $\sigma_\mathrm{RM}-\tau$ (p-value of 0.024: $\sim2\sigma$) in the FRB rest frame (Figure \ref{fig:sigmarm_correlation}). Note that, in this work, we consider a p-value smaller than $0.05$ (2$\sigma$) to be marginally significant, and less than $0.003$ (3$\sigma$) to be significant.

\subsection{Magnetized host environment} \label{sec:results_locenv}
The observed DM and RM are the sum of individual contributions from plasma in distinct environments along the LOS,
\begin{multline}
\mathrm{DM} = \mathrm{DM}_{\mathrm{ion}} + \mathrm{DM}_{\mathrm{disk}} + \mathrm{DM}_{\mathrm{halo}} + \mathrm{DM}_{\mathrm{IGM}}(z) \\
+ \frac{\mathrm{DM}_{\mathrm{host}}}{(1 + z)}~\mathrm{pc}~\mathrm{cm}^{-3}\,; \label{eq:dm_comps}
\end{multline}
\begin{multline}
\mathrm{RM} = \mathrm{RM}_{\mathrm{ion}} + \mathrm{RM}_{\mathrm{disk}} + \mathrm{RM}_{\mathrm{halo}} + \mathrm{RM}_{\mathrm{IGM}}(z) \\
+ \frac{\mathrm{RM}_{\mathrm{host}}}{(1+z)^2}~\mathrm{rad}~\mathrm{m}^{-2}\,, \label{eq:rm_comps}
\end{multline}
where ``ion" refers to the Earth's ionosphere, ``disk" and ``halo", are the Milky Way disk and halo, respectively, ``IGM" is the intergalactic medium and ``host" is the host galaxy. With $\mathrm{DM}_{\mathrm{host}}$ and $\mathrm{RM}_{\mathrm{host}}$, we can calculate the electron density-weighted average LOS magnetic field strength in the FRB host galaxy as:
\begin{equation}
\left<B_{\parallel,\mathrm{host}}\right> = 1.232 \frac{\mathrm{RM}_\mathrm{host}}{\mathrm{DM}_\mathrm{host}}~\mu\text{G}\,. \label{eq:b_host}
\end{equation}

Following \cite{2025ApJ...982..146P}, we estimate $\mathrm{DM}_\mathrm{host}$, $\mathrm{RM}_\mathrm{host}$, and $\left<B_{\parallel,\mathrm{host}}\right>$ and their respective uncertainties using Monte Carlo random sampling. For each of $10^4$ trials, we draw all foreground DM and RM components from their assumed distributions and use the host redshift $z=0.1839 \pm 0.0001$ \citep{2026Sci...391..280L} to compute rest-frame quantities. We then take the median and $68$\% confidence intervals as the maximum likelihood estimate and uncertainties. We adopt the same assumed foreground component distributions as \cite{2025ApJ...982..146P}:
\begin{enumerate}
    \item \textit{Observed DM and RM:} drawn from a Gaussian distribution with mean equal to the measured DM/RM and the standard deviation equal to the measurement uncertainty.
    \item \textit{Ionosphere:} considered to be negligible since the ionosphere typically only contributes $\sim 10^{-5}~\mathrm{pc}~\mathrm{cm}^{-3}$ to the DM \citep{2016ApJ...821...66L} and $\sim 0.1-1~\mathrm{rad}~\mathrm{m}^{-2}$ \citep{2019MNRAS.484.3646S} to the RM.
    \item \textit{Galactic disk:} $\mathrm{DM}_\mathrm{disk}$ is taken as the mean value estimated by the NE2001 model \citep[$40~\mathrm{pc}~\mathrm{cm}^{-3}$;][]{2002astro.ph..7156C, Ocker_2024_RNAAS}, with associated 20\% uniform uncertainty. The Milky Way RM contribution and its associated uncertainty is estimated using the \cite{2022A&A...657A..43H} reconstructed Galactic RM mean and variance maps at the FRB position, $\mathrm{RM}_\mathrm{disk} + \mathrm{RM}_\mathrm{halo} = -53 \pm 7~\mathrm{rad}~\mathrm{m}^{-2}$.
    \item \textit{Galactic halo:} $\mathrm{DM}_\mathrm{halo}$ is drawn from a Lognormal distribution with mean $\mathrm{log}_{10}(30~\mathrm{pc}~\mathrm{cm}^{-3})$ and a 0.2~dex standard deviation \citep{2015MNRAS.451.4277D,2020ApJ...888..105Y,2023ApJ...946...58C}. We have accounted for $\mathrm{RM}_\mathrm{halo}$ above.
    \item \textit{IGM:} $\mathrm{DM}_\mathrm{IGM}$ is sampled directly from the probability distribution function presented in Equation 1 of \cite{2024ApJ...965...57B}, which follows the empirical ``Macquart relation'' \citep{2020Natur.581..391M}. $\mathrm{RM}_\mathrm{IGM}$ is sampled from a Gaussian distribution with mean $0~\mathrm{rad}~\mathrm{m}^{-2}$ and standard deviation $6~\mathrm{rad}~\mathrm{m}^{-2}$ \citep{2010MNRAS.409L..99S}.
\end{enumerate}

In the CHIME/FRB data, we find that FRB~20220529A's $\mathrm{DM}_\mathrm{host}$ varies between $80^{+30}_{-41}~\mathrm{pc}~\mathrm{cm}^{-3}$ at MJD~59746 and $77^{+30}_{-41}~\mathrm{pc}~\mathrm{cm}^{-3}$ at MJD~60898 and the $\mathrm{RM}_\mathrm{host}$ varies between $+68 \pm 13~\mathrm{rad}~\mathrm{m}^{-2}$ and $+140 \pm 13~\mathrm{rad}~\mathrm{m}^{-2}$. The uncertainties on $\mathrm{DM}_\mathrm{host}$ and $\mathrm{RM}_\mathrm{host}$ are dominated by the uncertainties in the foreground DM and RM estimates; they do not reflect the precision with which we measure variations in the observed DM and RM over time. We implicitly assume that the foreground DM and RM contributions do not vary on month-to-year timescales \citep{2013MNRAS.435.1610P, 2017ApJ...841..125J}. Under this assumption, the observed variations map directly to changes in $\mathrm{DM}_\mathrm{host}$ and $\mathrm{RM}_\mathrm{host}$, and therefore originate locally to FRB~20220929A. Taking our $\mathrm{DM}_\mathrm{host}$ estimate as an upper limit for the DM contributed locally to FRB~20220529A shows that the integrated electron column in its source environment has decreased by $\geq 3.5\pm0.2$\% over $3.2$~years ($\geq 1.10\pm0.05$\% per year). 

We compute the average LOS magnetic field strength in the host galaxy using Equation \ref{eq:b_host}, obtaining $1.1^{+1.2}_{-0.4}~\mu\mathrm{G} < \left<B_{\parallel,\mathrm{host}}\right> < 2.2^{+2.4}_{-0.6}~\mu\mathrm{G}$, directed towards us. 

\subsection{Magnetic field strength during the DM/RM excursion} \label{sec:b_flare}
We now explore the magnetoionic properties of the FRB~20220529A environment during the interval of temporally linked DM and RM increases. Their temporal association requires an increase in some combination of $n_\mathrm{e}$, $B_\parallel$, and/or the plasma path length local to the source. Following \cite{2023ApJ...951...82M}, we estimate the local LOS magnetic field using 
\begin{equation}
\left<B_{\parallel,\mathrm{local}}\right> = 1.232 \frac{\Delta\mathrm{RM}}{\Delta\mathrm{DM}}(1+z)~\mu\text{G}\,, \label{eq:b_local}
\end{equation}
where $\Delta\mathrm{DM}$ and $\Delta\mathrm{RM}$ are the total observed change in DM and RM during their respective peaks, and the $1+z$ factor converts to the rest frame. Assuming peak values of $\mathrm{DM} = 245.67 \pm 0.04~\mathrm{pc}~\mathrm{cm}^{-3}$ (this work) and $\mathrm{RM} = 1977 \pm 84~\mathrm{rad}~\mathrm{m}^{-2}$ \citep{2026Sci...391..280L}, we adopt $\Delta\mathrm{DM} = 0.85\pm0.06~\mathrm{pc}~\mathrm{cm}^{-3}$ (between MJDs 60243 and 60279), and $\Delta\mathrm{RM} = 1981 \pm 84~\mathrm{rad}~\mathrm{m}^{-2}$ (between our $\mathrm{MJD}~60243$ measurement and the FAST $\mathrm{MJD}~60292$ measurement), yielding $\left<B_{\parallel,\mathrm{local}}\right> = 3.4 \pm 0.2~\mathrm{mG}$.

This estimate of $\left<B_{\parallel,\mathrm{local}}\right>$ is subject to several caveats. First, we assume that the DM and RM increases observed between $\mathrm{MJDs}~60243 - 60292$ arise from the same physical medium local to the FRB source, which is supported by their close temporal association: the DM rise precedes the RM rise by only $13$~days, shorter than the observed timescale over which the RM declines to its original value, and the absence of other comparable DM/RM excursions observed in the CHIME/FRB or FAST data. Second, our measured $\Delta\mathrm{DM}$ and $\Delta\mathrm{RM}$ are likely lower limits since our sparse temporal sampling limits our ability to constrain the true peak DM and RM during the excursion. Hence, our quoted uncertainty of $\pm 0.2$~mG reflects only measurement errors and does not factor in the unknown true peak DM/RM amplitudes. Finally, we adopt our CHIME/FRB DM measurement from MJD~60279 as representative of the peak DM, since FAST DM measurements are more sensitive to burst morphology degeneracies (see Section~\ref{sec:results_dm}). Accordingly, $\left<B_{\parallel,\mathrm{local}}\right>$ should be interpreted as an order-of-magnitude estimate of the magnetic field local to FRB~20220529A during this interval.

Note that we cannot apply Equation \ref{eq:b_local} across the entire 3.2~years of DM and RM measurements since the long term DM and RM evolution of FRB~20220529A likely originate from distinct physical processes, as evidenced by the DM steadily declining while the RM undergoes multiple fluctuations and sign changes. We discuss this further in Section \ref{sec:discussion_dm_loss}.

\subsection{Search for an associated PRS} \label{sec:results_prs}
We searched for an associated PRS coincident with FRB~20220529A in archival VLA data (Section~\ref{sec:obs_vla}). We produced a cleaned image of the combined visibilities with rms noise $11~\mu\mathrm{Jy/beam}$. No source was detected at the FRB position, yielding a $5\sigma$ upper limit of $55~\mu\mathrm{Jy}$ at $1.5$~GHz. We convert this flux limit to a spectral luminosity using
\begin{equation}
L_\nu = \frac{4 \pi D_L^2 S_\nu}{(1+z)^{1+\beta}} \left( \frac{1.5~\mathrm{GHz}}{\nu} \right)^\beta\,, \label{eq:prs_lum}
\end{equation}
where $\beta$ is the radio spectral index, $D_L$ is the FRB luminosity distance, $S_\nu$ is the flux at frequency $\nu$, and $z=0.1839\pm0.0001$. Assuming $\beta = -0.4$ \citep{2017Natur.541...58C, 2022Natur.606..873N}, we constrain $L_\nu \leq 5 \times 10^{28}~\mathrm{erg}~\mathrm{s}^{-1}~\mathrm{Hz}^{-1}$ at $1.5$~GHz.

\section{Discussion} \label{sec:discussion}
Our results provide insight into FRB~20220529A's environment. We observe a multi-year DM decline interrupted by a brief rise occurring 13~days before a previously reported RM excursion, after which the DM resumes its decline. The inferred host galaxy magnetic field is modest ($1.1-2.2~\mu$G), but increases to $3.4 \pm 0.2$~mG during the DM/RM excursion, placing FRB~20220529A among the small number of FRBs with inferred $\sim$~mG-level local magnetic fields. For example, estimates of the local magnetic field strength around FRB~20121102A range from $\gtrsim 0.6-2.4$~mG \citep {2018Natur.553..182M} up to $17$~mG \citep{2021MNRAS.501L..76K}; comparable magnetic field strengths have also been inferred for FRBs~20190520B \citep[$3-6$~mG;][]{2023Sci...380..599A}, 20190417A \citep[$\gtrsim 0.65$~mG;][]{2025arXiv250905174M}, and 20240619D \citep[$0.27$~mG;][]{2025arXiv250916374O}. We also find scattering timescale variations on week-to-year timescales, evidence for depolarization between $1.25$~GHz and $600$~MHz, and place stringent upper limits on any associated PRS. Here, we synthesize these results to propose a physical scenario for the FRB~20220529A progenitor that can account for these observations.

\subsection{Sustained DM decline}  \label{sec:discussion_dm_loss}
The observed DM decline must be local to the FRB~20220529A environment, as our LOS through other foregrounds (Milky Way interstellar medium, intergalactic medium, and host galaxy interstellar medium) does not change appreciably over timescales of years \citep[e.g.,][]{2013MNRAS.435.1610P, 2017ApJ...841..125J}, and the ionosphere only causes small \citep[$\sim10^{-5}-10^{-4}~\mathrm{pc}~\mathrm{cm}^{-3}$;][]{2016ApJ...821...66L} stochastic DM variations. 

Long-term DM monitoring of Galactic pulsars provides useful context for FRB~20220529A. Millisecond ``spider" pulsars show DM variations as they pass behind companion material, but with amplitudes ($\sim10^{-3}-10^{-2}~\mathrm{pc}~\mathrm{cm}^{-3}$) far smaller than observed here \citep{2016MNRAS.462.1029S, 2020MNRAS.494.2948P, 2024ApJ...969...62S}. Larger DM variations of a few~$\mathrm{pc}~\mathrm{cm}^{-3}$ occur in pulsars orbiting massive companion stars (e.g., B1259$-$63 and J2108$+$4516), but these are short-lived changes and are phase locked with the orbital motion, unlike the sustained DM decline of FRB~20220529A \citep{2004MNRAS.351..599W, 2005MNRAS.358.1069J, 2023ApJ...943...57A}. Secular DM trends in isolated pulsars, attributed to motion through the interstellar medium, are approximately linear at times but orders of magnitude smaller ($\sim 10^{-3}~\mathrm{pc}~\mathrm{cm}^{-3}$; \citealt{2013ApJ...762...94D, 2013MNRAS.429.2161K,1997A&A...323..211C,2014ApJ...787...82F,2016ApJ...821...66L}). DM variations of $10^{-2}-10^{-1}~\mathrm{pc}~\mathrm{cm}^{-3}$ in the Crab pulsar arise from plasma structures in its nebula and can persist for weeks to months, but do not match the multi-year steady DM decline of FRB~20220529A \citep[e.g.,][]{2008A&A...483...13K, 2018MNRAS.479.4216M}. Notably, the Crab also exhibits scattering variability on comparable timescales, similar to that observed for FRB~20220529A and FRB~20190520B \citep{2023MNRAS.519..821O}, proposed to arise from discrete patches in the source environment \citep[possibly filamentary structure in the Crab's pulsar wind nebula;][]{1975MNRAS.172...97L, 2000ApJ...543..740B, 2001MNRAS.321...67L, 2018MNRAS.479.4216M}. 

Only three additional FRB sources have observed years-long DM decreases (FRBs~20121102A, 20180301A, and 20190520B) and an increasing DM has been seen in just two sources (in earlier observations of FRB~20121102A and in FRB~20240619D). The DM of FRB~20121102A first rose $1-3~\mathrm{pc}~\mathrm{cm}^{-3}$ over four years \citep{2019ApJ...876L..23H} before it decreased at a rate of $-4.5~\mathrm{pc}~\mathrm{cm}^{-3}~\mathrm{year}^{-1}$ over the subsequent three years \citep{2025arXiv250715790W, 2025arXiv251011352S}. FRB~20180301A decreased at $-2.7~\mathrm{pc}~\mathrm{cm}^{-3}~\mathrm{year}^{-1}$ over two years \citep[][]{2023MNRAS.526.3652K}, and the DM of FRB~20190520B fell at a rate of $-11~\mathrm{pc}~\mathrm{cm}^{-3}~\mathrm{year}^{-1}$ \citep{2025arXiv251207140W} to $-12.4~\mathrm{pc}~\mathrm{cm}^{-3}~\mathrm{year}^{-1}$ \citep{NIU202676} over four years. FRB~20240619D showed a moderate DM rise of $\sim 0.4~\mathrm{pc}~\mathrm{cm}^{-3}$ ($+2.4~\mathrm{pc}~\mathrm{cm}^{-3}~\mathrm{year}^{-1}$) over only a two-month interval \citep{2025arXiv250916374O}.

A leading explanation for gradually declining FRB DMs is an expanding supernova remnant around the progenitor \citep[e.g.,][]{2016MNRAS.458L..19C, 2016MNRAS.461.1498M, 2017ApJ...841...14M, 2017ApJ...841L..30P, 2017ApJ...847...22Y, 2018ApJ...861..150P}. \cite{2018ApJ...861..150P} build upon earlier work to account for a variable ionization fraction as the supernova remnant evolves; they evaluate the DM versus time ($t$) evolution in two scenarios: (i) a supernova remnant expanding into a uniform-density environment ($\mathrm{DM} \propto t^{-1/2}$ for the first $\sim 300-5000$~years post supernova, after which the DM begins to increase due to the contribution from the swept up interstellar material) and (ii) a supernova remnant expanding into an inhomogeneous stellar wind environment ($\mathrm{DM} \propto t^{-3/2}$ for $\gtrsim 10^3$~years after the supernova). Note that there could be added complexity in the $\mathrm{DM}-t$ relationship due to reionization of the recombined supernova ejecta, which is not considered in these models. The gradual, approximately linear DM decline over year-long timescales seen in FRB~20220529A is consistent with either of these scenarios.

In scenario (i), \cite{2018ApJ...861..150P} parameterize the rate of change in the DM contributed by the supernova remnant during the declining-DM phase as
\begin{multline}
\frac{\mathrm{dDM}_\mathrm{SNR}}{\mathrm{d}t} =  -26.4~\mathrm{pc}~\mathrm{cm}^{-3}~\mathrm{year}^{-1} \left(\frac{\mu}{\mu_e}\right)\\
\times \left( \frac{E_\mathrm{SN}}{10^{51}~\mathrm{erg}} \right)^{-1/4} \left( \frac{M_\mathrm{ej}}{1~\mathrm{M}_\odot} \right)^{3/4} \left( \frac{n_0}{1~\mathrm{cm}^{-3}} \right)^{1/2}\\
\times \left( \frac{t_\mathrm{SNR}}{1~\mathrm{year}} \right)^{-3/2}\,, \label{eq:dm_snr_cd}
\end{multline}
where $\mu$ and $\mu_e$ are the mean molecular weight and mean molecular weight per electron, respectively, $E_\mathrm{SN}$ is the kinetic energy of the initial supernova explosion, $M_\mathrm{ej}$ is the mass ejected by the stellar progenitor during the supernova, $n_0$ is the number density in the ambient environment, and $t_\mathrm{SNR}$ is the age of the supernova remnant. Using our rest-frame DM decline of $-1.235~\mathrm{pc}~\mathrm{cm}^{-3}~\mathrm{year}^{-1}$, and assuming the same typical values used by \cite{2018ApJ...861..150P} in their modeling (i.e., $\mu/\mu_e = 1$, $E_\mathrm{SN} = 10^{51}~\mathrm{erg}$, and $10^{-2}~\mathrm{cm}^{-3} \leq n_0 \leq 10^2~\mathrm{cm}^{-3}$), we determine $t_\mathrm{SNR}$ for $0.5~M_\odot < M_\mathrm{ej} < 15~M_\odot$, which spans the typical ranges expected for hydrogen-rich core-collapse \citep[i.e., Type II, $8~M_\odot < M_\mathrm{ej} < 15~M_\odot$;][]{2022A&A...660A..41M} and stripped-envelope \citep[i.e., Type Ib/c, $0.5~M_\odot < M_\mathrm{ej} < 8.3~M_\odot$;][]{2023ApJ...955...71R} supernovae. We derive the age of a supernova remnant expanding into a uniform-density environment associated with FRB~20220529A under these assumptions to be $1.2~\mathrm{years} \lesssim t_\mathrm{SNR} \lesssim 140~\mathrm{years}$. Our FRB observation baseline (3.2~years) is strictly inconsistent with the lower bound of this age range, thus disfavoring the combination of a small ejected mass ($M_\mathrm{ej} \sim 0.5~M_\odot$) and a low-density ($n_0 \sim 10^{-2}~\mathrm{cm}^{-3}$) ambient environment in scenario (i).

The DM rate of change contributed by a supernova remnant expanding into a stellar wind environment, i.e., scenario (ii), as described by \cite{2018ApJ...861..150P}, is
\begin{multline}
\frac{\mathrm{dDM}_\mathrm{SNR}}{\mathrm{d}t} =  -1.95\times10^4~\mathrm{pc}~\mathrm{cm}^{-3}~\mathrm{year}^{-1} \mu_e^{-1}\\
\times \left( \frac{E_\mathrm{SN}}{10^{51}~\mathrm{erg}} \right)^{-3/4} \left( \frac{M_\mathrm{ej}}{1~\mathrm{M}_\odot} \right)^{5/4} \left( \frac{K_\mathrm{wind}}{10^{13}~\mathrm{g}~\mathrm{cm}^{-1}} \right)^{1/2} \\
\times \left( \frac{t_\mathrm{SNR}}{1~\mathrm{year}} \right)^{-5/2}\,, \label{eq:dm_snr_pg}
\end{multline}
where $K_\mathrm{wind} = \dot{M}/4\pi v_\mathrm{wind}$ is the mass loading parameter for a stellar wind with mass loss rate $\dot{M}$ and velocity $v_\mathrm{wind}$. For the same $\mathrm{dDM}_\mathrm{SNR}/\mathrm{d}t$ and $M_\mathrm{ej}$ ranges as above, and assuming the typical values invoked by \cite{2018ApJ...861..150P} (i.e., $\mu_e = 1$, $E_\mathrm{SN} = 10^{51}~\mathrm{erg}$, and $10^{11}~\mathrm{g}~\mathrm{cm}^{-1} \leq K_\mathrm{wind} \leq 10^{15}~\mathrm{g}~\mathrm{cm}^{-1}$), we determine $14~\mathrm{years} \lesssim t_\mathrm{SNR} \lesssim 470~\mathrm{years}$ for a supernova expanding into a stellar wind environment.

Our supernova remnant age ranges are consistent with the $t_\mathrm{SNR} \lesssim 100$~year age estimate of FRB~20190520B assuming it is undergoing a similar power-law DM decline over time \citep{NIU202676}. The absence of any historical supernovae coincident with FRB~20220529A in the Transient Name Server \citep[e.g., as done by][]{2025ApJ...991..199D} may, in part, disfavor the youngest ages ($\sim$~years to a few decades) inferred in the two scenarios. 

In both scenarios (i) and (ii) above, we also expect a corresponding decrease in $|\mathrm{RM}|$ with time from the expanding supernova remnant. Using our Equation \ref{eq:b_local}, we expect a decrease in the $|\mathrm{RM}|$ of $\sim 2~\mathrm{rad}~\mathrm{m}^{-2}$ over 3.2~years, given a decrease in the DM of $2.7~\mathrm{pc}~\mathrm{cm}^{-3}$, and assuming a constant LOS magnetic field strength of $\sim 1~\mu\mathrm{G}$ (based on our $\left<B_{\parallel,\mathrm{host}}\right>$ estimate; Section \ref{sec:results_locenv}). However, FRB~20220529A shows both increases and decreases in its RM of a few $10^2~\mathrm{rad}~\mathrm{m}^{-2}$ over timescales of days to months in the FAST dataset, even outside of the RM excursion \citep[see our Figure 2 and][]{2026Sci...391..280L}. These RM variations are likely driven by rapid changes in the local magnetic field orientation projected along the LOS, which also explains the multiple RM sign changes near MJD~59800 over $\sim$~weeks. Due to these RM variations driven by a dynamic magnetic field orientation, and a dearth of polarized burst detections in the last $1-2$~years by both CHIME/FRB and FAST, we cannot rule out that the underlying baseline $|\mathrm{RM}|$ has declined in proportion to the observed DM decrease, as would be expected from an expanding supernova remnant.

\subsection{A local depolarizing environment} \label{sec:discussion_depol}
We measure spectral depolarization in FRB~20220529A across $400-1500$~MHz, characterized by an RM scatter of $\sigma_\mathrm{RM,host} = 3.50\pm0.03~\mathrm{rad}~\mathrm{m}^{-2}$ (rest frame; see Appendix \ref{a:depol}). For RM scatters of this magnitude, depolarization is expected to arise from inhomogeneous magnetoionic plasma near the FRB source \citep{2022Sci...375.1266F}. While correlations between $\sigma_\mathrm{RM,host}-|\mathrm{RM}_\mathrm{host}|$ and $\sigma_\mathrm{RM,host}-\tau_\mathrm{host}$ are only marginally significant across the current depolarizing FRB population (Appendix~\ref{a:depol}), this behavior could be explained by the Faraday rotation, scattering, and depolarization originating from the same local media in the FRB host galaxy. Our interpretation of FRB~20220529A being embedded within a supernova remnant, introduces a dynamic and inhomogeneous environment which can naturally explain the observed depolarization.

Variance in the $L/I$ versus frequency may be attributed to a combination of years-long temporal variations in the RM scatter \citep[as observed in FRB~20180916B;][]{2024MNRAS.527.9872G}, and variation in the intrinsic $L/I$ of bursts. Thus, the $\sigma_\mathrm{RM}$ estimates for all repeating FRBs, including FRB~20220529A, should be considered average values weighted by the burst activity of the source.

\subsection{Limits on associated PRS emission} \label{sec:discussion_prs}
Four repeating FRBs have been discovered to be co-located with a compact ($\lesssim 1-10~\mathrm{pc}$) PRS \citep{2017ApJ...834L...8M, Bhandari_2023_ApJL, 2025arXiv250905174M, Bruni_2025_A&A}. Leading theories to explain this continuum radio emission often invoke a young magnetar central engine that both produces the FRBs and drives a highly magnetized wind nebula that produces the PRS through interactions with the supernova remnant \citep{Beloborodov17, Margalit18, Li20, Bhattacharya24}. 

As we have discussed, the multi-year DM decline, depolarization, and temporal RM and scattering variability of FRB~20220529A are consistent with such a supernova remnant environment, motivating our search for associated PRS emission. We place an upper limit of $L_\nu \leq 5\times10^{28}~\mathrm{erg~s^{-1}~Hz^{-1}}$ at 1.5~GHz (Section~\ref{sec:results_prs}), which is lower than the PRS luminosities of FRBs~20121102A, 20190529B, and 20190417A by a factor of $\sim2-6$, and comparable to that of FRB~20240114A. The PRS luminosity predicted from the $L_\nu$–$|\mathrm{RM}|$ relation by \cite{2020ApJ...895....7Y} for nebular emission is $L_\nu\sim10^{27}~\mathrm{erg~s^{-1}~Hz^{-1}}$, well below our observational limit, and we therefore cannot rule out the presence of a faint PRS consistent with this framework. 

\subsection{What caused the DM/RM excursion?} \label{sec:discussion_dm_rise}
Equipped with some insights into the properties and long-term evolution of the FRB~20220529A environment, we now examine various physical scenarios that may explain the short-lived rise and fall seen in the source's DM and RM.

\subsubsection{Interaction with a young supernova remnant} \label{sec:discussion_sn_interaction}
Given our interpretation that FRB~20220529A originates from within a supernova remnant, we first explore scenarios in which the DM/RM excursions can be produced in such an environment. 

\cite{2026Sci...391..280L} argue against a supernova remnant plasma overdensity causing the RM excursion, based on differences in the temporal RM structure function during the excursion versus the FRB's ``normal'' state. This inherently assumes that all structures in the supernova remnant arise from the same turbulent power-law scaling, but it is possible that some plasma structures could deviate from this. A dense plasma clump crossing the FRB sightline for $27$~days (the period of heightened DM/RM) with a transverse velocity of $10 - 1000~\mathrm{km}~\mathrm{s}^{-1}$ \citep[similar to filament velocities in the Crab Nebula;][]{1968AJ.....73..535T, 2008MNRAS.384.1200R} would have a size of $\sim 0.16-16$~AU. Assuming the clump has a magnetic field of $B\sim 3.4$~mG and contributes $\sim 2000~\mathrm{rad}~\mathrm{m}^{-2}$ to the RM \citep{2026Sci...391..280L}, the electron density of the clump is $9\times10^{3}~\mathrm{cm}^{-3}\lesssim n_\mathrm{e} \lesssim 9\times10^{5}~\mathrm{cm}^{-3}$, consistent with the density of some filaments in young supernova remnants \citep[e.g. G11.2$-$0.3 and 0540$‑$69.3;][]{2013ApJ...770..143L, 2025MNRAS.542.2830T}. With a clump electron density of $\sim 9\times10^{3}~\mathrm{cm}^{-3} - 9\times10^{5}~\mathrm{cm}^{-3}$ and size $\sim 0.16-16$~AU, the estimated DM contribution is $\sim 0.7~\mathrm{pc}~\mathrm{cm}^{-3}$, comparable to our observed DM rise, suggesting that an extremely dense, magnetized plasma clump could explain the short-lived DM/RM jump.

Alternatively, if the FRB source is a magnetar, the magnetar ejecta that propagate out at a high velocity could interact with the supernova remnant material creating a region of shocked, ionized gas along the FRB LOS, contributing to the DM and RM. Models of colliding magnetar shells at different velocities reproduce the RM excursion peak and subsequent decline over $\sim$weeks \citep{2025A&A...698L...3X}, and relativistic ejecta can increase the DM by $\sim 1~\mathrm{pc}~\mathrm{cm}^{-3}$ for days-to-months \citep{2019MNRAS.485.4091M}, consistent with our observations of FRB~20220529A. While \cite{2026Sci...391..280L} argue against this scenario due to the absence of a contemporaneous burst rate increase with the RM rise, it is possible that the magnetar flaring activity precedes the RM peak, with a delay required for the ejecta to reach the supernova remnant material. For example, ejecta launched at $0.7c$ \citep[e.g., as measured for an SGR~1806-20 flare;][]{2006ApJ...638..391G} around MJD~60000 (when FAST observed a relatively high FRB rate) could traverse $\sim 0.18$~pc in 300~days before interacting with the supernova remnant material and causing a DM/RM change. A young supernova remnant with radius of order $0.1$~pc is consistent with the model proposed by \cite{2018ApJ...861..150P} given our observed DM decline. While out of the scope of this work, hydrodynamic simulations of magnetar flare ejecta interactions in a supernova environment would help model the observed DM/RM evolution.

\subsubsection{CME from a binary companion} \label{sec:discussion_cme}
\cite{2026Sci...391..280L} proposed that the RM excursion in FRB~20220529A most likely arises from a CME launched by a binary stellar companion intersecting the FRB LOS. They consider two scenarios: (i) a magnetized M dwarf CME with mass $\sim 10^{17}$~g, surface magnetic field $\sim 10^4$~G, and an initial CME size $\sim10^{-1}~\mathrm{R}_\odot$, or (ii) a giant star or Algol binary CME \citep[e.g.,][]{2017ApJ...850..191M} with mass $\sim 10^{21}$~g, surface magnetic field $\sim 1$~G, and initial CME size $\sim 10^{-1}~\mathrm{R}_\odot$. The M dwarf model predicts a DM contribution of only $\sim0.01~\mathrm{pc}~\mathrm{cm}^{-3}$, inconsistent with our observed $\sim 1~\mathrm{pc}~\mathrm{cm}^{-3}$ DM increase 13~days before the RM excursion. In contrast, the giant star/Algol DM contribution prediction ($\sim 10~\mathrm{pc}~\mathrm{cm}^{-3}$) is consistent with our data, given that we place only a lower limit on the peak DM.

Again, following the assumptions in \cite{2026Sci...391..280L}, i.e. assuming the magnetic field scaling with radial distance, $B \propto r^{-2}$: our inferred LOS field of $\sim 3.4$~mG implies a distance of $\sim 17~\mathrm{R}_\odot$ from the companion. For CME velocities of $10^2 - 10^3~\mathrm{km}~\mathrm{s}^{-1}$, this corresponds to a delay of $\sim 0.1 - 1$~days between CME launch and FRB sightline crossing, one to three orders of magnitude shorter the best-fit delays constrained by \cite{2026Sci...391..280L}. Even adopting a $10~\mathrm{pc}~\mathrm{cm}^{-3}$ DM increase, as predicted for the giant star/Algol scenario, and the observed RM peak (yielding $B_{\parallel}\sim0.3$~mG), we obtain $\sim58~\mathrm{R}_\odot$ and a delay of $\sim0.5$–$5$ days, still shorter than the best-fit CME model \citep{2026Sci...391..280L}. This disagreement could be alleviated by assuming a shallower magnetic-field profile \citep[e.g., $B \propto r^{-1.57}$;][]{2025arXiv251204730M}, and/or a larger initial CME magnetic field strength. We therefore cannot exclude a giant star/Algol CME origin for the DM and RM excursion. However, explaining the long-term DM decline would additionally require a secularly decreasing electron column, such as from an expanding supernova remnant surrounding the system.

\subsubsection{Stellar winds or accretion from a massive binary companion} \label{sec:discussion_binary}
If binary systems such as pulsars B1259$-$63 and J2108$+$4516, which show large DM/RM increases during periastron, were embedded in an expanding supernova remnant, they may be able to reproduce both the long-term and short-lived DM/RM variability in FRB~20220529A. \cite{2026Sci...391..280L} argue this scenario is unlikely to explain the RM excursion based on the extreme eccentricities and orbital periods required. We revisit this model using our 3.2~year dataset. The effective duty cycles for increase DM/RM in pulsars B1259$-$63 and J2108$+$4516 are $\sim 0.08$ and $\sim 0.4$, respectively \citep{2005MNRAS.358.1069J, 2023ApJ...943...57A}. For FRB~20220529A, the DM/RM excursion spans MJDs~60279 (the DM rise we observe) to 60306 \citep[when the RM returns to $\lesssim 100~\mathrm{rad}~\mathrm{m}^{-2}$][]{2026Sci...391..280L}, giving a duty cycle $\lesssim 0.023$ assuming the period is longer than our observational campaign. We caution that low burst activity of FRB~20220529A, as observed by CHIME/FRB, over the last $\sim1.5$~years may have resulted in us missing additional DM/RM excursions. Following \cite{2026Sci...391..280L} (see their Section S2.1.3), this implies a binary eccentricity $e_\mathrm{orb} \gtrsim 0.94$, consistent with their results. Rare, long-period, highly eccentric pulsars exist \citep[e.g., pulsars J1638$-$4725 and J2032$+$4127;][]{2006MNRAS.372..777L, 2015MNRAS.451..581L}, so this remains a viable explanation for the DM/RM excursion. Such a system would need to be embedded in an environment, such as an expanding supernova remnant, to additionally explain the long-term DM decline.

\section{Conclusions} \label{sec:conclusions}
We report the discovery and $\sim3.2$~year monitoring of FRB~20220529A with CHIME/FRB, detecting $16$ bursts above S/N of $12$, of which $7$ are significantly polarized. Our low frequency observations ($400-800$~MHz) allow high-precision tracking of frequency-dependent propagation effects (DM, RM, and scattering) and comparison with $1.0-1.5$~GHz observations by \cite{2026Sci...391..280L}. We additionally use archival VLA data to place stringent upper limits on an associated PRS and derive physical properties of the source environment. Our key conclusions from this work are:
\begin{enumerate}
    \item \textbf{Long-term DM decline.} We discover a steady decrease in the observed DM over multiple years at a rate of $-0.881\pm0.001~\mathrm{pc}~\mathrm{cm}^{-3}~\mathrm{year}^{-1}$ ($-1.235\pm0.001~\mathrm{pc}~\mathrm{cm}^{-3}~\mathrm{year}^{-1}$ rest frame), corresponding to a $\geq 3.5\pm0.2$\% decrease in the electron column of the FRB~20220529A source environment. This is consistent with the FRB being embedded in a $\sim$~years to centuries old expanding supernova remnant \citep{2018ApJ...861..150P}, suggesting a young compact object from a massive stellar progenitor as the FRB source. 
    \item \textbf{Magnetoionic variability decoupled from secular DM evolution.} While the steady DM decline is consistent with an expanding supernova remnant, the $|\mathrm{RM}|$ does not show the corresponding smooth decrease of $\sim 2~\mathrm{rad}~\mathrm{m}^{-2}$ expected over the same interval. Instead, we observe RM variations on the order of $10$~rad~m$^{-2}$, and \cite{2026Sci...391..280L} report variations of $\sim10^2$~rad~m$^{-2}$ on days-to-month timescales, including sign reversals. These fluctuations point to rapid changes in the local magnetic field orientation. We also observe scattering timescale variations of $\sim1$~ms at 600~MHz on similar timescales, comparable to the Crab pulsar and consistent with discrete plasma structures in the source environment. Additionally, we detect spectral depolarization ($\sigma_{\rm RM,host}=3.50\pm0.03~\mathrm{rad,m^{-2}}$), indicative of inhomogeneous magnetoionic plasma near the source. Together, these results imply a dynamic source environment, consistent with FRB~20220529A residing within a supernova remnant with filamentary structures and a complex magnetic field configuration.
    \item \textbf{Short-lived DM/RM excursion.} We detect a DM increase, deviating from the multi-year DM decline, $\sim13$~days before the reported RM excursion \citep{2026Sci...391..280L}. During this interval of heightened DM and RM, the LOS magnetic field strength reaches $\left<B_{\parallel,\mathrm{local}}\right> = 3.4 \pm 0.2~\mathrm{mG}$, among the highest measured for an FRB. A dense plasma structure within a young supernova remnant can account for such a DM/RM enhancement, consistent with Conclusions 1 and 2. Alternative scenarios include a CME or wind interaction from a binary companion, or magnetar ejecta propagating outward and shocking the surrounding remnant.
\end{enumerate}
Continued monitoring of FRB~20220529A is paramount for determining whether the DM decline persists or if the source environment enters a new evolutionary phase. For example, \cite{2018ApJ...861..150P} predict that a supernova remnant expanding into a uniform-density medium will exhibit a rising DM in later evolutionary stages, whereas expansion into a stellar wind environment should continue to produce a declining DM for $\gtrsim 10^3$~years. Long-term observations will also test whether additional RM/DM excursion episodes occur, as expected in some binary interaction models \citep[][and references therein]{2026Sci...391..280L}.

\section*{Acknowledgements}
We acknowledge that CHIME is located on the traditional, ancestral, and unceded territory of the Syilx/Okanagan people. We are grateful to the staff of the Dominion Radio Astrophysical Observatory, which is operated by the National Research Council of Canada. CHIME is funded by a grant from the Canada Foundation for Innovation (CFI) 2012 Leading Edge Fund (Project 31170) and by contributions from the provinces of British Columbia, Québec and Ontario. The CHIME/FRB Project is funded by a grant from the CFI 2015 Innovation Fund (Project 33213) and by contributions from the provinces of British Columbia and Québec, and by the Dunlap Institute for Astronomy and Astrophysics at the University of Toronto. Additional support was provided by the Canadian Institute for Advanced Research (CIFAR), McGill University and the Trottier Space Institute at McGill thanks to the Trottier Family Foundation, and the University of British Columbia. The baseband recording system for CHIME/FRB is funded in part by a CFI John R. Evans Leaders Fund award to IHS. The AstroFlash research group at McGill University, University of Amsterdam, ASTRON, and JIVE is supported by: a Canada Excellence Research Chair in Transient Astrophysics (CERC-2022-00009); an Advanced Grant from the European Research Council (ERC) under the European Union's Horizon 2020 research and innovation programme (`EuroFlash'; Grant agreement No. 101098079); an NWO-Vici grant (`AstroFlash'; VI.C.192.045); an ERC Starting Grant (`EnviroFlash'; Grant agreement No. 101223057); and an NWO-Veni grant (VI.Veni.222.295). This work is based in part on observations carried out using the 32-m radio telescope operated by the Institute of Astronomy of the Nicolaus Copernicus University in Toru\'n (Poland) and supported by a Polish Ministry of Science and Higher Education SpUB grant. We acknowledge the use of public data from the DESI Legacy Survey. This work makes use of data from the Westerbork Synthesis Radio Telescope, owned by ASTRON. ASTRON, the Netherlands Institute for Radio Astronomy, is an institute of the Dutch Scientific Research Council NWO (Nederlandse Oranisatie voor Weten-schappelijk Onderzoek). We thank the Westerbork operators Richard Blaauw, Jurjen Sluman and Henk Mulder for scheduling and supporting observations.

A.P. is a Trottier Space Institute Postdoctoral Fellow.
K.N. acknowledges support by NASA through the NASA Hubble Fellowship grant \# HST-HF2-51582.001-A awarded by the Space Telescope Science Institute, which is operated by the Association of Universities for Research in Astronomy, Incorporated, under NASA contract NAS5-26555.
A.M.C. is a Banting Postdoctoral Researcher.
A.P.C. is a Canadian SKA Scientist and is funded by the Government of Canada/est financé par le gouvernement du Canada.
V.M.K. holds the Lorne Trottier Chair in Astrophysics \& Cosmology, a Distinguished James McGill Professorship, receives support from an NSERC Discovery grant (RGPIN 228738-13), and thanks Tel Aviv University for their hospitality.
M.L. acknowledges the support of the Natural Sciences and Engineering Research Council of Canada (NSERC-CGSD)
K.W.M. is supported by NSF Grant Nos. 2008031, 2510771 and holds the Adam J. Burgasser Chair in Astrophysics.
D.M. acknowledges support from the French government under the France 2030 investment plan, as part of the Initiative d'Excellence d'Aix-Marseille Universit\'e -- A*MIDEX (AMX-23-CEI-088).
M.N. is a Fonds de Recherche du Quebec – Nature et Technologies (FRQNT) postdoctoral fellow.
A.B.P. acknowledges support by NASA through the NASA Hubble Fellowship grant HST-HF2-51584.001-A awarded by the Space Telescope Science Institute, which is operated by the Association of Universities for Research in Astronomy, Inc., under NASA contract NAS5-26555. A.B.P. also acknowledges prior support from a Banting Fellowship, a McGill Space Institute~(MSI) Fellowship, and a Fonds de Recherche du Quebec -- Nature et Technologies~(FRQNT) Postdoctoral Fellowship.
M.W.S is a Fonds de Recherche du Quebec - Nature et Technologies (FRQNT) postdoctoral fellow and acknowledges support from the Trottier Space Institute Fellowship program.
P.S. acknowledges the support of an NSERC Discovery Grant (RGPIN-2024-06266).
V.S. is supported by a Fonds de Recherche du Quebec - Nature et Technologies (FRQNT) Doctoral Research Award.

\facilities{CHIME, RT-1 Westerbork, Toru\'n, and VLA.}

\software{{\tt Astropy} \citep{2013A&A...558A..33A, 2018AJ....156..123A, 2022ApJ...935..167A},
{\tt DM\_PHASE} \citep{2019ascl.soft10004S},
{\tt fitburst} \citep{2024ApJS..271...49F},
{\tt Matplotlib} \citep{Hunter:2007}, 
{\tt NumPy} \citep{harris2020array},
{\tt PyGEDM} \citep{2021PASA...38...38P},
{\tt RM-CLEAN} \citep{2009A&A...503..409H},
{\tt RM-synthesis} \citep{2005A&A...441.1217B}, 
{\tt RM-tools} \citep{2020ascl.soft05003P, 2026arXiv260120092V}, and
{\tt SciPy} \citep{2020SciPy-NMeth}.}

\appendix

\section{CHIME/FRB Outrigger localization} \label{sec:appendix_A}
Of the 16 bursts detected by CHIME/FRB from FRB~20220529A, five were recorded in voltage data at at least one of the CHIME/FRB Outrigger telescopes \citep{2025ApJ...993...55C}. These data enable cross-correlation between CHIME and its Outriggers to obtain precise burst localizations using very long baseline interferometry (VLBI). 

Bursts 10 and 11 were recorded only at the k'ni\textipa{P}atn k'l$\left._\mathrm{\smile}\right.$stk'masqt (KKO) Outrigger, corresponding to a CHIME-KKO baseline of $66$~km \citep{Lanman_2024_AJ}. Theoretically, such a baseline provides arcsecond-level localization precision along a single axis, with the perpendicular constraint supplied by the CHIME baseband localization \citep{2021ApJ...910..147M}. Bursts 12 and 13 were recorded at both KKO and the Green Bank Outrigger (GBO), introducing a longer baseline of $\sim3300$~km, theoretically improving the localization precision to $50$~milliarcseconds (mas). Burst 16 was recorded at all three Outrigger sites which, in theory, should result in a localization precision of $50\times100$~mas. 

We formed cross-correlated visibilities between CHIME and each Outrigger site using {\tt pyfx} \citep{Leung_2025_AJ}, detecting VLBI fringes for all five bursts on each baseline for which voltage data were available. Instrumental delays were corrected using in-beam steady source calibration \citep{Andrew_2025_ApJ} using calibrator sources in the Radio Fundamental Catalog \citep{Petrov_2025_ApJS} observable in the FRB baseband dump, listed in Table~\ref{tb:loc_props}. We note that although GBO voltage data were recorded for burst 13, and VLBI fringes from the target were significantly detected, the in-beam calibrator was not successfully detected on the CHIME-GBO baseline. This is most likely due to differences in the effective field of view, which was shifted $\sim 1^\circ$ in hour angle between burst 12 and 13, moving the calibrator from $\sim1^\circ$ to $\sim2^\circ$ relative to the beam center. For burst 12, the calibrator J0119$+$3210 is detected with a coherent S/N of 13.4 across the frequency range where the FRB is bright, marginally exceeding our minimum threshold of S/N$=13$ for a reliable calibrator solution. For burst 13, however, the increased angular separation from the beam center leads to a primary beam response decrease of more than 50\% at $450$~MHz (where the burst is detected), reducing the calibrator sensitivity to a coherent S/N of $\sim6$, well below our threshold. Consequently, for burst 13 we are limited to a CHIME-KKO localization only. After in-beam calibration, we fit out any residual ionospheric delay contribution (which has a $1/\nu$ spectral behavior in the visibility phases; see Equation A2 in \citealt{Leung_2025_AJ}).

The localization independently measured for each burst are shown in Table~\ref{tb:loc_props}, where our uncertainties are inflated compared to the theoretical expectations listed above. Here, the localization uncertainties are dominated by systematic effects, characterized through test localizations (using the VLBI positions of pulsars determined through Very Long Baseline Array campaigns: project codes VLBA/21A-314, VLBA/22A-345 and VLBA/23A-099; Curtin et al. {\it in prep}), and dependent upon the burst S/N, the bandwidth of the emission, and the target-calibrator separation (e.g. similar behavior was observed using the European VLBI Network in \citealt{2021arXiv211101600N}). Typically, the target-calibrator separation has a minor impact if the burst is bright and broadband, however, the narrow emitting bandwidth and relatively low S/N of the bursts in our sample (Figure~\ref{fig:waterfalls}) results in comparatively large uncertainties, especially since our calibrator-target separations are relatively large ($>10^{\circ}$; Table~\ref{tb:loc_props}). Further, while burst 16 is detected across the full CHIME/FRB Outrigger array, the S/N is too low to achieve theoretical expectations for our uncertainties. 

\begin{figure}
    \centering
    \includegraphics[width=0.528\columnwidth]{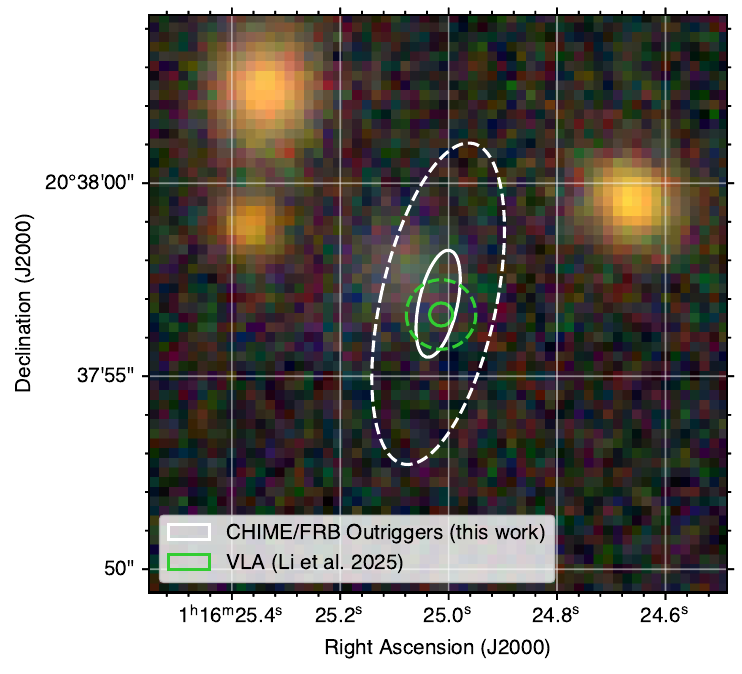}
    \caption{FRB~20220529A sky position as measured by the CHIME/FRB Outriggers in this work (white), corresponding to the combined localization of bursts 12 and 16, compared to the position reported by the VLA (green; \citealt{2026Sci...391..280L}). Solid and dashed ellipses represent the $1\sigma$ and $3\sigma$ localization uncertainties, respectively. The localization ellipses are overplotted on an archival Dark Energy Camera Legacy Survey $gri$-filter image of the field. The faint host galaxy identified by \cite{2026Sci...391..280L} can be seen consistent with both localization ellipses.}
    \label{fig:loc}
\end{figure}

Within our uncertainties we find all of our bursts to be consistent in their spatial location with each other. Further, to improve on the individual burst localizations, we compute the combined localization, where we treat the localizations of bursts 12 and 16 as statistically independent and multiply their corresponding localization probability distributions (Figure~\ref{fig:loc}). Note that we exclude bursts 10, 11 and 13, which do not affect the combined localization due to the large uncertainties reported from the single CHIME-KKO baseline. We measure the combined position of FRB~20220529A to be Right Ascension (ICRS) $={01{\mathrm{h}}16{\mathrm{m}}24.99{\mathrm{s}}}$ and Declination (ICRS) $={20^{\circ}37^{\prime}56.55^{\prime\prime}}$, with localization uncertainties characterized by a semi-major axis $a_{\rm err}=1.39^{\prime\prime}$, semi-minor axis $b_{\rm err}=0.64^{\prime\prime}$, and position angle $\theta=-12.6^\circ$ East of North. This is consistent with the independent VLA/{\it realfast} localization reported by \citet{2026Sci...391..280L}. A comparison of the CHIME/FRB Outrigger localization and the VLA localization, overlaid on an archival Dark Energy Camera Legacy Survey $gri$-band image of the field \citep{2019AJ....157..168D}, is shown in Figure~\ref{fig:loc}. 

\setlength{\tabcolsep}{0.55pt}
\setlength\extrarowheight{-1pt}
\setlength{\LTcapwidth}{1.0\textwidth}
\begin{table*}[ht!]
\begin{center}
\caption{Summary of localization properties for FRB~20220529A bursts detected and localizated by the CHIME/FRB Outriggers.} \label{tb:loc_props}
\begin{tabular}{ccccccccc}
\hline
\hline
Burst & Outrigger sites & Right Ascension & Declination & $b_{\rm err}$  & $a_{\rm err}$ & $\theta$ & In-beam & Target-calibrator\\
& & (ICRS) & (ICRS) & ($\prime\prime$) &  ($\prime\prime$) &  ($^\circ$)  & calibrator & separation ($^\circ$)\\
\hline
10 & KKO & ${01{\mathrm{h}}16{\mathrm{m}}25.00{\mathrm{s}}}$ & ${20^{\circ}37^{\prime}48.14^{\prime\prime}}$ & $1.98$ & $22.5$ & $8.9$ & J0117$+$8928 & $68.8$\\
11 & KKO & ${01{\mathrm{h}}16{\mathrm{m}}24.84{\mathrm{s}}}$ & ${20^{\circ}37^{\prime}39.03^{\prime\prime}}$  & $2.01$ & $31.2$ & $8.2$ & J0119$+$0829 & $12.1$\\
12 & KKO, GBO & ${01{\mathrm{h}}16{\mathrm{m}}25.87{\mathrm{s}}}$ & ${20^{\circ}38^{\prime}53.06^{\prime\prime}}$ & $0.63$ & $31.3$ & $12.0$ & J0117$+$4536 (KKO), J0119$+$3210 (GBO) & $24.9$, $11.6$\\
13 & KKO, GBO$^{a}$ & ${01{\mathrm{h}}16{\mathrm{m}}24.99{\mathrm{s}}}$ & ${20^{\circ}37^{\prime}47.74^{\prime\prime}}$ & 2.13 & 22.7 & 9.1 & J0108$+$0135 & 19.1\\
16 & KKO, GBO, HCO &  ${01{\mathrm{h}}16{\mathrm{m}}25.06{\mathrm{s}}}$ & ${20^{\circ}37^{\prime}56.38^{\prime\prime}}$ & $1.41$ & $1.99$ & $101.5$ & J0119$+$3210 & $11.6$ \\
\hline
\hline
\end{tabular}
\end{center}

\tablenotetext{$a$}{We were unable to detect an in-beam calibrator at GBO for burst 13, so this data was not used for the burst localization.} 
\end{table*}

\section{Hyperflash detection burst properties} \label{sec:appendix_B}
We detected a single burst with Westerbork at L-band and report a ToA of MJD~$60924.00396628$. This MJD is referenced with respect to infinite frequency in the dynamical barycenter timescale, see \cite{2025arXiv250916374O} for the full method. We optimized the DM by sampling the peak S/N across a range of DM values at a time resolution of $512$~$\upmu$s. These DM values were taken between $242$ and $244.5~\mathrm{pc}~\mathrm{cm}^{-3}$, with a step size of $0.01~\mathrm{pc}~\mathrm{cm}^{-3}$ (250 steps). Each DM data product was made with the Super~FX~Correlator ({\tt SFXC}), which enables coherent (within channels) and incoherent (between channels) dedispersion \citep{keimpema_2015_exa}. We find an optimized DM value for the Westerbork burst of: DM$_\textrm{Wb} = 243.21\pm0.41~\mathrm{pc}~\mathrm{cm}^{-3}$ (left panel of Appendix Figure~\ref{fig:westerbork_burst_and_dm}). We quote $1\sigma$ errors on the DM, defined as the drop of the fitted Gaussian peak S/N by $1$. The fluence of the burst is measured by applying the radiometer equation over the entire bandwidth of the burst, and is found to be $10.9 \pm 2.2$~Jy~ms, assuming a System Equivalent Flux Density (SEFD) of 420~Jy\footnote{\url{https://www.evlbi.org/sites/default/files/shared/EVNstatus.txt}}. Due to the limited dynamic range of 2-bit sampling at Westerbork, we experience saturation effects for bright bursts and can correct for these when needed \citep[e.g.,][]{ouldboukattine_2025_mnras}. In this case, the burst is not bright enough for saturation to significantly affect the measured energy, and we therefore did not apply a saturation correction.

\begin{figure}
    \centering
    \includegraphics[width=0.49\columnwidth]{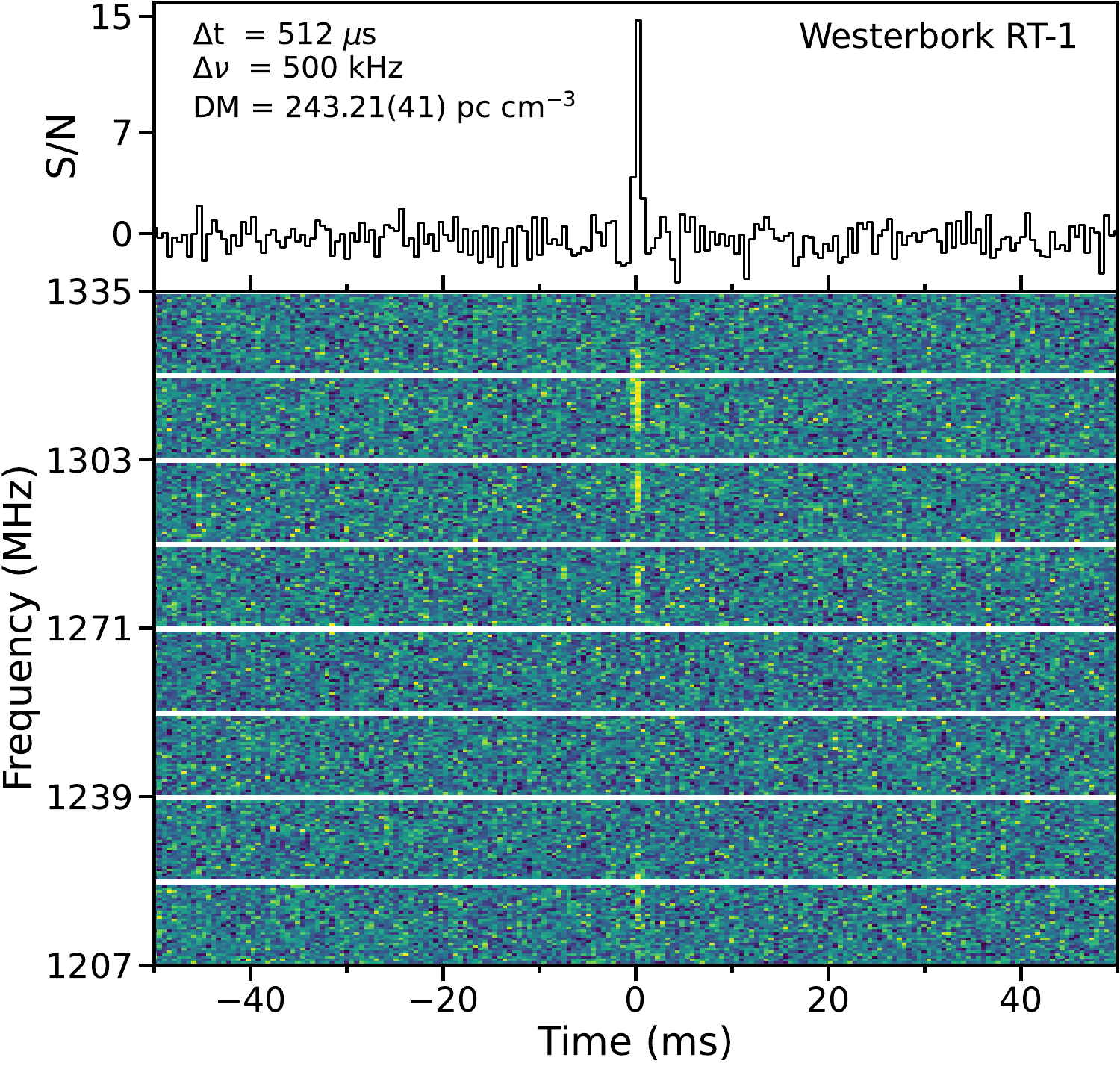}
    \includegraphics[width=0.468\columnwidth]{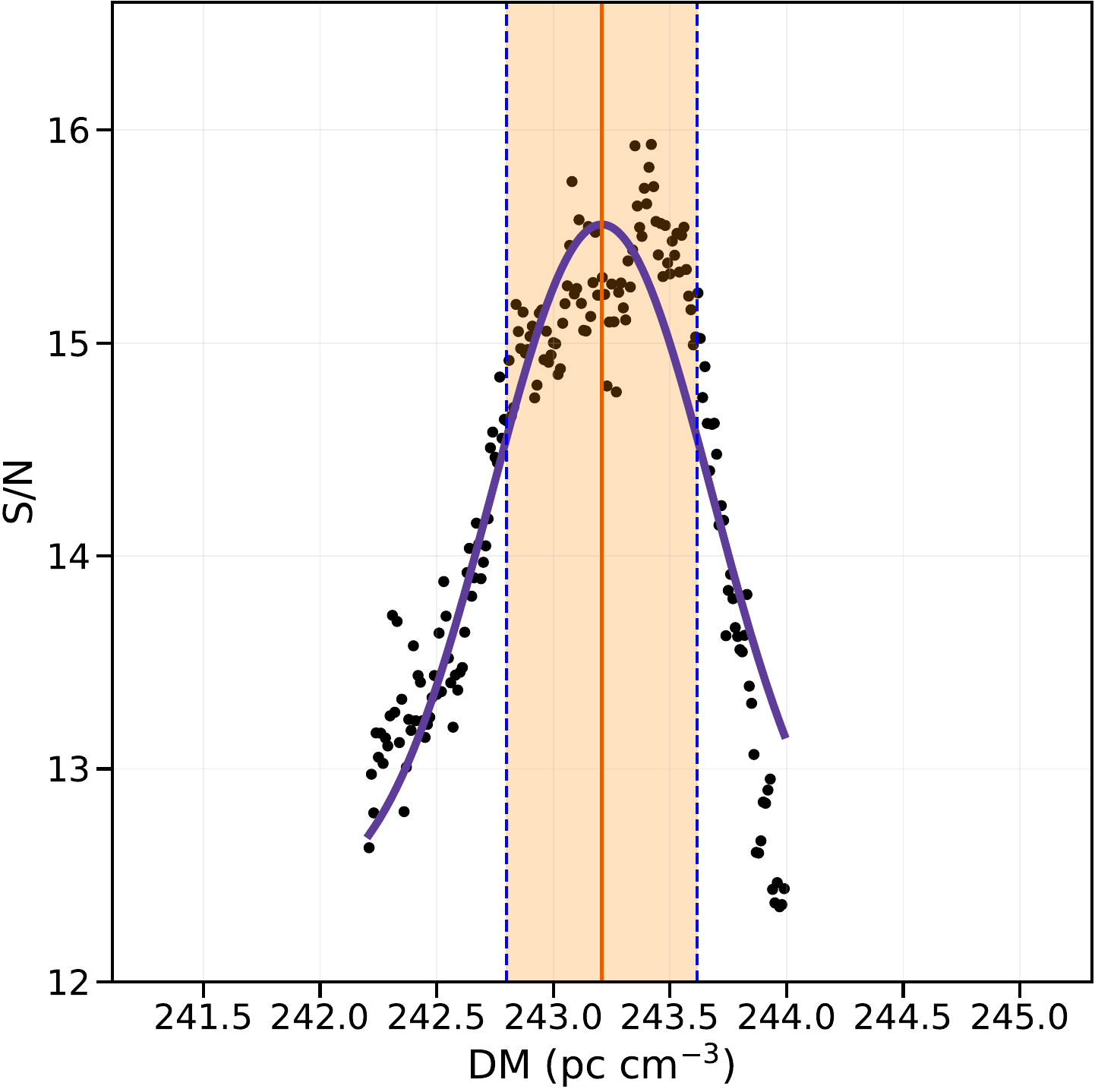}
    \caption{(Left) L-band dynamic spectrum of the FRB~20220529A burst detected by Westerbork. A frequency-averaged burst profile is presented in the top panel. (Right) Peak S/N of the burst at a range of DM trials with the peak DM represented by a red vertical line and the $1\sigma$ uncertainty region around the peak, where the S/N drops by 1, is highlighted in orange.}
    \label{fig:westerbork_burst_and_dm}
\end{figure}

\section{Correlation between depolarization, Faraday rotation, and scattering}\label{a:depol}
A correlation between $\sigma_\mathrm{RM}$ and $|\mathrm{RM}|$ was first suggested by \cite{2022Sci...375.1266F} for seven repeating FRBs, and \cite{2025MNRAS.tmp.1894U} later evaluated this relation with the addition of one non-repeating FRB. Both of their analyses used the observer frame $\sigma_\mathrm{RM}$ and RM instead of the corrected rest frame values, which can lead to a bias in the relationship, especially given the limited number of FRBs in which depolarization has been observed. We convert to the rest-frame RM scatter, $\sigma_\mathrm{RM,host}$, using the known redshifts of the host galaxies of these depolarizing FRBs \citep[see][and references therein]{2022Sci...375.1266F, 2025MNRAS.tmp.1894U},
\begin{equation}
\sigma_\mathrm{RM,host} = \sigma_\mathrm{RM} (1+z)^2\,, \label{eq:sigma_rm_host}
\end{equation}
\citep[e.g., see][]{2024MNRAS.528.2511T}. This gives $\sigma_\mathrm{RM,host} = 3.50\pm0.03~\mathrm{rad}~\mathrm{m}^{-2}$ for FRB~20220529A. We also take the following steps for the RMs of FRB~20220529A and the other depolarizing FRBs reported by \cite{2022Sci...375.1266F} and \cite{2025MNRAS.tmp.1894U}:
\begin{enumerate}
    \item We subtract the Galactic RM, as measured by \cite{2022A&A...657A..43H}, from each polarized burst and multiply the resulting RM by $(1+z)^2$ to get the corresponding rest frame $\mathrm{RM}_\mathrm{host}$ (assuming $\mathrm{RM}_\mathrm{IGM} = 0~\mathrm{rad}~\mathrm{m}^{-2}$).
    \item We take the absolute value of $\mathrm{RM}_\mathrm{host}$ for all polarized burst(s) and then, if it is a repeater, we take the median $\mathrm{RM}_\mathrm{host}$ across those polarized bursts.
\end{enumerate}
The first step is important for sources for which the amplitude of the Galactic RM contribution is comparable to, or larger than, the observed RM (e.g., FRB~20220529A), and for sources at high $z$. The second step is crucial for repeaters that undergo changes in their RM sign over time \citep[e.g., FRB~20190520B;][]{2023Sci...380..599A}, as averaging before taking the absolute value could lead to a much lower $|\mathrm{RM}_\mathrm{host}|$. 

\begin{figure*}
    \centering
    \includegraphics[width=0.49\textwidth]{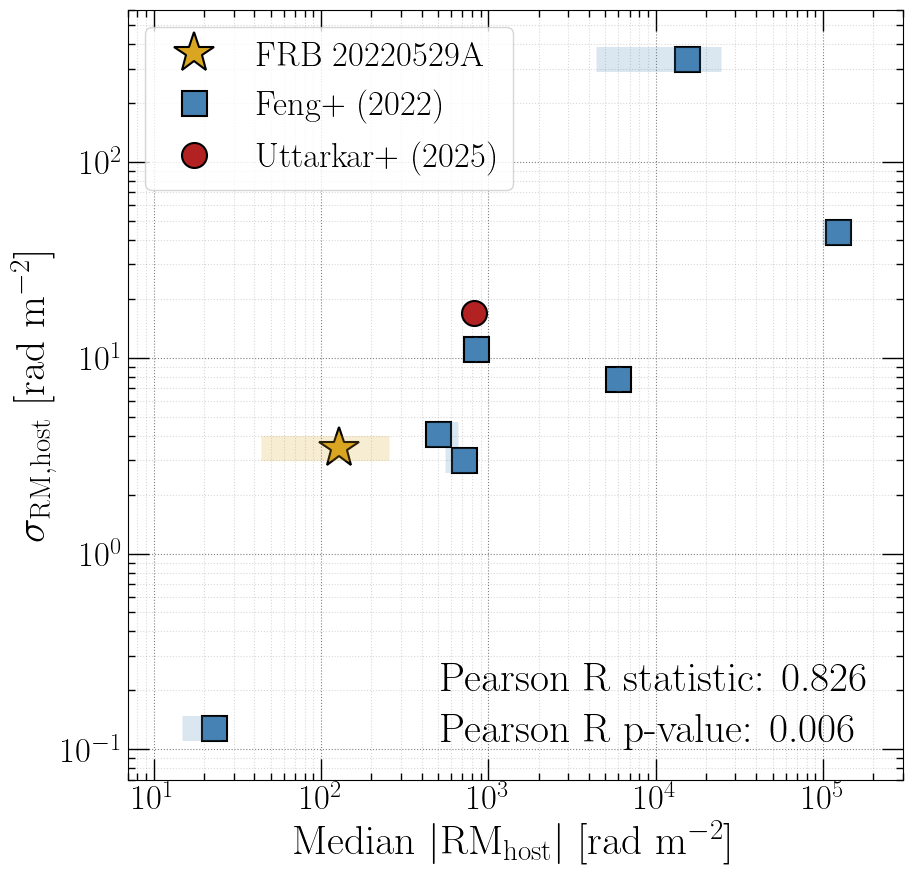}
    \includegraphics[width=0.49\textwidth]{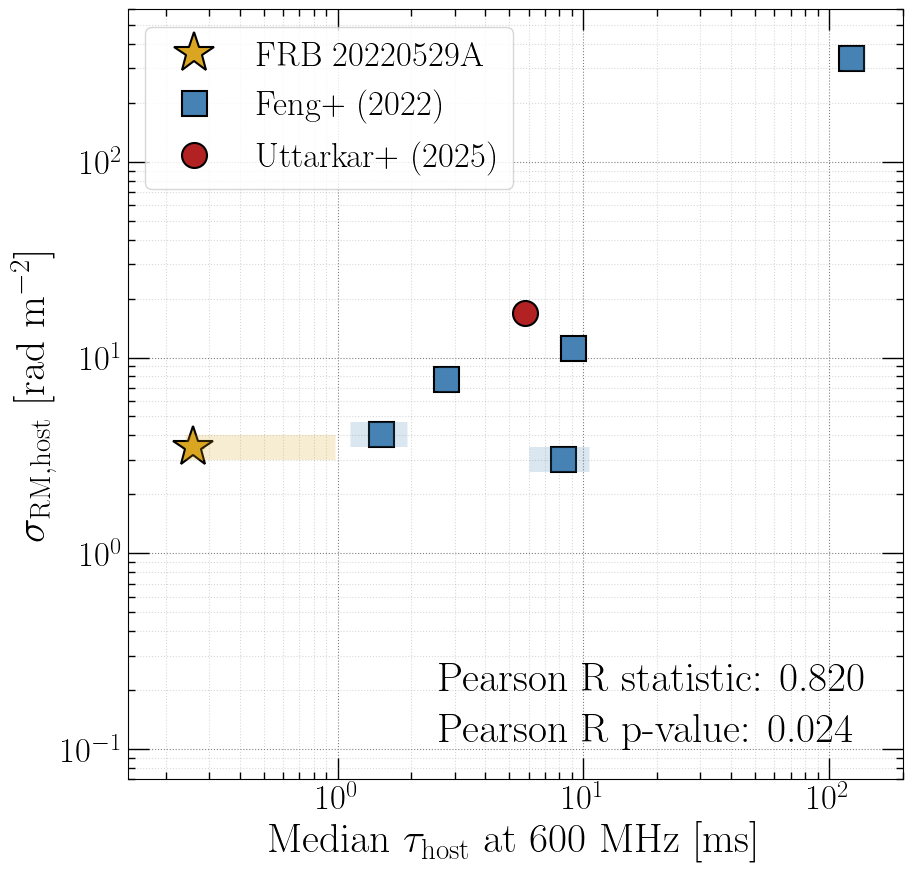}
    \caption{(Left) Rest-frame scatter in the observed RM, $\sigma_\mathrm{RM,host}$, plotted as a function of median $|\mathrm{RM}_\mathrm{host}|$ for FRB~20220529A and for other FRBs from the literature. (Right) Median rest-frame scattering timescale referenced to 600~MHz, where available, versus $\sigma_\mathrm{RM,host}$ for the same FRBs. Gold and blue shaded regions represent the 68\% confidence interval around the median $|\mathrm{RM}_\mathrm{host}|$ and $\tau$ for FRB~20220529A and other repeaters, respectively. Uncertainties on $\sigma_\mathrm{RM,host}$ are smaller than the markers. The test statistics and p-values from Pearson R tests in both cases are presented in the bottom right of each panel.} 
    \label{fig:sigmarm_correlation}
\end{figure*}

In the left panel of Figure \ref{fig:sigmarm_correlation}, we plot the median $|\mathrm{RM}_\mathrm{host}|$, i.e., averaged across all polarized CHIME/FRB and FAST bursts outside of the RM excursion, versus $\sigma_\mathrm{RM,host}$ of FRB~20220529A (gold star) and compare it to other FRBs in which depolarization has been measured: \cite{2022Sci...375.1266F} (all repeaters; blue squares) and \cite{2025MNRAS.tmp.1894U} (apparent non-repeater; red circle). For the repeating FRBs, we plot the 68\% confidence interval around the median $|\mathrm{RM}_\mathrm{host}|$ as shaded regions. We conduct a Pearson R test on the $\mathrm{log}_{10}(\sigma_\mathrm{RM,host})-\mathrm{log}_{10}(|\mathrm{RM}_\mathrm{host}|)$ relationship and we find a positive correlation (test statistic equal to 0.826) with a p-value equal to 0.006. Thus, this $\mathrm{log}_{10}(\sigma_\mathrm{RM,host})-\mathrm{log}_{10}(|\mathrm{RM}_\mathrm{host}|)$ relationship is marginally significant.

Next, we evaluate the relationship between observed depolarization and scattering in FRB~20220529A and other depolarizing FRBs from the literature. We restrict our sample to sources for which the measured scattering timescale exceeds the Milky Way contribution predicted by the NE2001 electron column density model \citep[scaled to the same reference frequency assuming a frequency dependence of $\nu^{-4}$;][]{2002astro.ph..7156C, Ocker_2024_RNAAS} by more than an order of magnitude \citep{2019MNRAS.486.3636P, 2022ApJ...927...59L, 2023MNRAS.519..821O, 2025ApJ...992..206C, 2025PASA...42..133S}. This selection allows us to assume that the observed broadening is likely originating from the FRB host galaxy or its local environment. Similar to the RMs, scattering contributions from multiple screens along the LOS add linearly \citep[e.g.][]{Cordes_2022_ApJ}. Following naturally from our selection criteria above, the predicted Galactic interstellar medium contribution is much less than the measured scattering timescales \citep{2002astro.ph..7156C, Ocker_2024_RNAAS}, and we further assume that any contribution from intervening galaxy halos is negligible. Under these assumptions, the measured scattering timescales, $\tau$, are dominated by scattering media local to the FRB source. 

To account for the fact that scattering occurs at an emitted frequency $\nu(1+z)$ and that the observed scattering timescale is subject to cosmological time dilation, we correct the measured scattering values to the rest frame by dividing by a factor $(1+z)^{3}$, assuming $\tau\propto\nu^{-4}$ \citep{Macquart_2013_ApJ}, such that
\begin{equation}
\tau_\mathrm{host} = \tau/(1+z)^3\,. \label{eq:tau_host}
\end{equation}
Note that if the scattering index diverges from the assumed $\nu^{-4}$, it can change the significance of any observed correlation. The rest frame scattering timescales, $\tau_\mathrm{host}$, scaled to a reference frequency of 600~MHz to facilitate comparison with our CHIME measurements of FRB~20220529A, are plotted versus $\sigma_\mathrm{RM,host}$ in the right panel of Figure \ref{fig:sigmarm_correlation}. The respective 68\% confidence intervals around the median $\tau_\mathrm{host}$ for repeaters are plotted as shaded regions. Using a Pearson R test on $\mathrm{log}_{10}(\sigma_\mathrm{RM,host})-\mathrm{log}_{10}(\tau_\mathrm{host})$ we find a marginally significant positive correlation with a p-value of 0.024 and a test statistic of $0.820$.

\bibliography{ref}{}
\bibliographystyle{aasjournalv7}

\end{document}